\documentclass{article}
\usepackage{fullpage}

\usepackage{bussproofs}
\usepackage{tikzfig}
\tikzstyle{every picture}=[baseline=-0.25em,scale=0.6]
\tikzstyle{dotpic}=[scale=0.6]
\tikzstyle{diredges}=[every to/.style={diredge}]
\tikzstyle{dot graph}=[shorten <=-0.1mm,shorten >=-0.1mm,scale=0.6]
\tikzstyle{digraph}=[-latex]
\tikzstyle{plot point}=[circle,fill=black,minimum width=2mm,inner sep=0]
\tikzstyle{string graph}=[scale=0.6]
\tikzstyle{sg diredge}=[-stealth]
\tikzstyle{rewrite edge}=[-open triangle 45]
\tikzstyle{sg bold diredge}=[-stealth,thick,shorten >=-1pt]
\tikzstyle{sg vertex}=[circle,minimum width=2.2mm,fill=white,draw=black,inner sep=0mm]
\tikzstyle{labelled sg vertex}=[circle,minimum width=7mm,fill=white,draw=black,inner sep=0mm]
\tikzstyle{sg grey vertex}=[sg vertex,fill=gray!30!white]
\tikzstyle{sg black vertex}=[sg vertex,fill=black]
\tikzstyle{sg bold vertex}=[circle,minimum width=2.2mm,fill=white,draw=black,very thick,inner sep=0mm]
\tikzstyle{sg wire vertex}=[circle,minimum width=1mm,fill=black,inner sep=0mm]

\tikzstyle{tick vertex}=[rectangle,fill=black,minimum height=1mm,minimum width=2.5mm,inner sep=0mm]

\tikzstyle{braceedge}=[decorate,decoration={brace,amplitude=2mm,raise=-1mm}]
\tikzstyle{small braceedge}=[decorate,decoration={brace,amplitude=1mm,raise=-1mm}]
\tikzstyle{left hook arrow}=[left hook-latex]
\tikzstyle{right hook arrow}=[right hook-latex]

\tikzstyle{dot}=[inner sep=0.7mm,minimum width=0pt,minimum height=0pt,fill=black,draw=black,shape=circle]
\tikzstyle{white dot}=[dot,fill=white]
\tikzstyle{alt white dot}=[white dot,label={[xshift=2.9mm,yshift=-0.1mm]left:$\cdot$}]
\tikzstyle{gray dot}=[dot,fill=gray!50]
\tikzstyle{box vertex}=[draw=black,rectangle]
\tikzstyle{whitebg}=[fill=white,inner sep=2pt]
\tikzstyle{graph state vertex}=[sg vertex,fill=black]

\tikzstyle{wide point}=[fill=white,draw=black,shape=isosceles triangle,shape border rotate=90,isosceles triangle stretches=true,inner sep=1pt,minimum width=1.5cm,minimum height=5mm]
\tikzstyle{wide copoint}=[fill=white,draw=black,shape=isosceles triangle,shape border rotate=-90,isosceles triangle stretches=true,inner sep=1pt,minimum width=1.5cm,minimum height=5mm]
\tikzstyle{symm}=[ultra thick,shorten <=-1mm,shorten >=-1mm]

\tikzstyle{square box}=[rectangle,fill=white,draw=black,minimum height=6mm,minimum width=6mm]
\tikzstyle{square gray box}=[rectangle,fill=gray!30,draw=black,minimum height=6mm,minimum width=6mm]
\tikzstyle{point}=[regular polygon,regular polygon sides=3,draw=black,scale=0.75,inner sep=-0.5pt,minimum width=7mm,fill=white]
\tikzstyle{copoint}=[point,regular polygon rotate=180,fill=white]
\tikzstyle{gray point}=[point,fill=gray!40!white]
\tikzstyle{gray copoint}=[copoint,fill=gray!40!white]

\tikzstyle{open graph}=[baseline=-0.25em]
\tikzstyle{greybg}=[background rectangle/.style={fill=black!5,draw=black!30,rounded corners=1ex}, show background rectangle]

\tikzstyle{edge point}=[circle,minimum width=1mm,fill=black,inner sep=0mm]
\tikzstyle{vertex point}=[circle,minimum width=2.2mm,fill=white,draw=black,inner sep=0mm]
\tikzstyle{gray vertex point}=[circle,minimum width=2.2mm,fill=gray!30!white,draw=black,inner sep=0mm]
\tikzstyle{edge label}=[inner sep=2pt, font=\small]
\tikzstyle{on edge label}=[fill=white, font=\footnotesize, inner sep=1 pt]

\newcommand{\edgearrow}{{\arrow[black]{>}}}
\newcommand{\edgetick}{{\arrow[black,scale=0.7,very thick]{|}}}
\tikzstyle{diredge}=[postaction=decorate,decoration={markings, mark=at position 0.55 with \edgearrow}]
\tikzstyle{medium diredge}=[postaction=decorate,decoration={markings, mark=at position 0.75 with \edgearrow}]
\tikzstyle{short diredge}=[->]
\tikzstyle{halfedge}=[-)]
\tikzstyle{other halfedge}=[(-]
\tikzstyle{freeedge}=[(-)]
\tikzstyle{white edge}=[line width=5pt,white]
\tikzstyle{tick}=[postaction=decorate,decoration={markings, mark=at position 0.5 with \edgetick}]
\tikzstyle{small map edge}=[|-latex, gray!60!blue, shorten <=0.9mm, shorten >=0.5mm]
\tikzstyle{thick dashed edge}=[very thick,dashed,gray!40]
\tikzstyle{dashed edge}=[densely dotted,thick]
\tikzstyle{map edge}=[|-latex,very thick, gray!40, shorten <=1mm, shorten >=0.5mm]
\tikzstyle{tickedge}=[postaction=decorate,
  decoration={markings, mark=at position 0.5 with \edgetick}]
\tikzstyle{dirtickedge}=[postaction=decorate,
  decoration={markings, mark=at position 0.5 with \edgetick},
  decoration={markings, mark=at position 0.85 with \edgearrow}]
\tikzstyle{dirdoubletickedge}=[postaction=decorate,
  decoration={markings, mark=at position 0.4 with \edgetick},
  decoration={markings, mark=at position 0.6 with \edgetick},
  decoration={markings, mark=at position 0.85 with \edgearrow}]

\tikzstyle{arrs}=[-latex,font=\small,auto]
\tikzstyle{arrow plain}=[arrs]
\tikzstyle{arrow dashed}=[dashed,arrs]
\tikzstyle{arrow bold}=[very thick,arrs]
\tikzstyle{arrow hide}=[draw=white!0,-]
\tikzstyle{arrow reverse}=[latex-]
\tikzstyle{cdnode}=[]

\tikzstyle{cnot}=[fill=white,shape=circle,inner sep=-1.4pt]
\tikzstyle{bang box}=[draw=black,dashed,minimum height=12mm,minimum width=12mm,fill=gray!20]
\tikzstyle{wire label}=[font=\footnotesize, auto]

\usepackage{noncommgraph}

\newcommand{\showproofs}{1}

\usepackage{amsmath}
\usepackage{amssymb}
\usepackage{stmaryrd}
\usepackage{amsthm}
\usepackage{xspace}
\usepackage{hyperref}
\usepackage{graphicx}
\graphicspath{{figures/}}

\usepackage{algpseudocode}
\usepackage{algorithm}
\algblock[Switch]{Switch}{EndSwitch}
\algblock[Case]{Case}{EndCase}
\algtext*{EndCase}

\newtheorem{theorem}{Theorem}[section]
\newtheorem*{theorem*}{Theorem}

\newtheorem{lemma}[theorem]{Lemma}
\newtheorem{corollary}[theorem]{Corollary}
\theoremstyle{definition}
\theoremstyle{definition}
\theoremstyle{definition}\newtheorem{definition}[theorem]{Definition}
\theoremstyle{definition}
\theoremstyle{definition}\newtheorem{remark}[theorem]{Remark}
\theoremstyle{definition}
\theoremstyle{definition}
\theoremstyle{definition}
\theoremstyle{definition}
\theoremstyle{definition}\newtheorem{notation}[theorem]{Notation}

\tikzstyle{cdiag}=[matrix of math nodes, row sep=3em, column sep=3em, text height=1.5ex, text depth=0.25ex,inner sep=0.5em]
\tikzstyle{arrow above}=[transform canvas={yshift=0.5ex}]
\tikzstyle{arrow below}=[transform canvas={yshift=-0.5ex}]

\def\bR{\begin{color}{red}} 
\def\bB{\begin{color}{blue}}
\def\bM{\begin{color}{magenta}}
\def\bC{\begin{color}{cyan}}
\def\bW{\begin{color}{white}}
\def\bBl{\begin{color}{black}} 
\def\bG{\begin{color}{green}}
\def\bY{\begin{color}{yellow}}
\def\bDG{\begin{color}{green!70!black}}
\def\e{\end{color}}

\renewcommand{\showproofs}{1}

\begin{document}

\title{Tensors, !-graphs, and non-commutative quantum structures}
\author{Aleks Kissinger\\
{\normalsize University of Oxford}\\
{\small\tt aleks.kissinger@cs.ox.ac.uk}
\and
David Quick\\
{\normalsize University of Oxford} \\
{\small\tt david.quick@cs.ox.ac.uk}}

\maketitle

\begin{abstract}
	!-graphs provide a means of reasoning about infinite families of string diagrams and have proven useful in manipulation of (co)algebraic structures like Hopf algebras, Frobenius algebras, and compositions thereof. However, they have previously been limited by an inability to express families of diagrams involving non-commutative structures which play a central role in algebraic quantum information and the theory of quantum groups. In this paper, we fix this shortcoming by offering a new semantics for non-commutative !-graphs using an enriched version of Penrose's abstract tensor notation.
\end{abstract}

\tableofcontents

\newpage


\section{Introduction}\label{sec:intro}

\textit{Diagrammatic theories} give us a way to study a wide variety of algebraic and coalgebraic structures in monoidal categories. They have played a big role in categorical quantum mechanics, almost since its inception~\cite{CD1,ContPhys,CDKZ}, and are becoming increasingly important in areas as disparate as computational linguistics~\cite{DimitriDPhil} and control theory~\cite{SobocinskiSignal,Baez2014a}.

A diagrammatic theory consists of two parts: a \textit{signature} $\Sigma$ and a set of \textit{diagram equations} $E$. The signature consists of a set of objects $\{ A, B, \ldots \}$ along with a set of generating morphisms with input and output arities formed from combining objects with $\otimes$ and $I$. For example, the signature of a Frobenius algebra consists of four morphisms: $(\mu : A \otimes A \to A,\ \eta : I \to A,\ \delta : A \to A \otimes A,\ \epsilon : A \to I)$, or, written diagrammatically:
\[ \Sigma \ =\  \input{frob_sig.tikz} \]
If necessary we could label the generators and/or edges by their names/types. In this case there is only one edge type and the generators are uniquely defined by their arities.

Then, $E$ is a set of equations between morphisms built from these generators, which we can picture as equations between string diagrams. For example, the theory of commutative Frobenius algebras contains the (co)associativity, (co)unit, (co)commutativity and Frobenius equations: \\
\begin{equation}\label{eqn:frob-laws}
	E := \left\{\quad\input{comm_frob_eqns.tikz}\quad\right\}
\end{equation}
A \textit{model} of $(\Sigma, E)$ in a (symmetric, traced, or compact closed) monoidal category $\mathcal C$ assigns a morphism to each generator in $\Sigma$ such that all equations in $E$ hold.

\begin{remark}
	Many familiar algebraic constructions arise as special cases of this setup. For instance, any linear `term-like' algebraic theory (i.e. where free variables occur precisely once on the LHS and RHS of every equation) can be presented this way. Also, if we restrict to equations in $E$ that are directed acyclic, we obtain presentations of PROPs (or coloured PROPs in the multi-sorted case). In that case, models of $(\Sigma, E)$ in $\mathcal C$ are in 1-to-1 correspondence with strong monoidal functors from the presented PROP into $\mathcal C$.
\end{remark}

This style of algebraic theory works well when generators have fixed, finite arity. However, it is often possible to find a much more elegant presentation of a theory if we allow the arity of our generators to vary. For instance, commutative Frobenius algebras can be alternatively presented using a single variable-arity generator sometimes called a `spider', along with just two equations.
\[
\Sigma \ =\ \left\{ \ \input{spider-node.tikz} \ \right\} \qquad
E \ =\ \left\{ \ \input{spider_merge.tikz}, \ \ \ \
                        \input{spider_elim.tikz} \ \right\}
\]
A model of such a theory is no longer just a finite set of morphisms, but rather, a set of \textit{families} of morphisms $f_{j,k} : A^{\otimes j} \to A^{\otimes k}$, indexed by input/output arities, such that the equations in $E$ hold for all possible arities.

Comparing this to the equations at the beginning of this section, we seem to have lost some formality. That is, the `concrete' diagrammatic identities above can be formalised in such a way that proofs can be performed (and even machine-checked) via a suitable notion of diagram rewriting, as formalised in~\cite{DixonKissinger2010}.
This level of rigour is lost when we describe equations in a mathematical meta-language, making use of ellipses, for example, to represent repetition. However, in~\cite{DixonDuncan2009}, the authors introduced \textit{!-boxes} (pronounced `bang-boxes') as a method for reasoning about graphs with repeated structure. As !-box rules, the previously informal rules can be formalised as:
\[
\Sigma \ =\ \left\{ \ \input{spider_bb.tikz} \ \right\} \qquad
E \ =\ \left\{ \ \input{spider_merge_bb.tikz}, \quad
                        \input{spider_elim.tikz} \ \right\}
\]
Intuitively, marking a subgraph with a !-box means that subgraph (along with edges in/out of it) can be repeated any number of times to obtain an \textit{instance} of the graph. Thus we interpret a graph with !-boxes as a set of all its instances.
\[ \left\llbracket \input{spider_bb.tikz} \quad \right\rrbracket
   \ :=\  \left\{ \ \ \input{spider_inst1.tikz} \ \ \right\} \]
Similarly, for rules with !-boxes, matched pairs of !-boxes can be repeated in the LHS and RHS to obtain instances of that rule.
\[ \left\llbracket \input{spider_merge_bb.tikz} \quad \right\rrbracket
   \ :=\  \left\{ \ \ \input{spider_merge_inst.tikz} \ \ \right\} \]

Thus, for our example of the commutative Frobenius algebra, we have reduced our theory of 7 equations to just 2.

!-boxes were given a formal semantics in~\cite{PatternGraph}, making use of adhesive categories~\cite{LackAdh2005}. Using these semantics, !-box rewriting has also been implemented in the graphical proof assistant Quantomatic~\cite{Quantomatic}. They also come with a simple and powerful induction principle introduced by one of the authors in~\cite{KissingerThesis} and proven correct in~\cite{MerryThesis}. But there's a catch: note how we were careful to say that \textit{commutative} Frobenius algebras have an elegant presentation as above. A major drawback of the existing !-box notation is that it is only unambiguous if all of the nodes in the diagram are invariant under permuting inputs/outputs. This is severely limiting in two ways. The first and most obvious limitation is that we are forced to consider only commutative algebraic structures. The second, more subtle limitation is that we have no freedom to \textit{definitionally} extend our theory, i.e. introduce new nodes defined as diagrams of other nodes, without making implicit assumptions about those diagrams (namely, that they are symmetric on inputs/outputs).

In order to overcome these shortcomings, we extend the !-graph notation with some extra information about how newly-created edges should be ordered when a !-box is expanded. This turns out to be fairly straightforward as soon as one shifts from a graph-based semantics for diagrams, as employed in~\cite{DixonKissinger2010}, to a \textit{tensor-based} semantics, where morphisms in the free compact closed category are represented using a version of Penrose's abstract tensor notation~\cite{Penrose1971}. This approach, recently formalised in~\cite{KissingerATS}, has the property that non-commutativity comes `for free', where the edges connected to a single element are represented as a list of edge names. This essentially accomplishes the same thing as equipping each vertex in a graph with a total ordering on its adjacent edges---which appears in Joyal and Street's original geometric construction~\cite{JS}---but has the advantage that this syntax can be annotated with precisely the extra data needed to expand !-boxes in an unambiguous way.



The paper is structured as follows. In Section~\ref{sec:tensors}, we introduce a tensor notation of compact closed categories, and show that tensor expressions can be used to characterise the free compact closed category. In Section~\ref{sec:bboxes}, we extend tensor expressions to include !-boxes. We furthermore introduce a graphical notation for these !-tensors, which we call (non-commutative) !-graph notation. In Section~\ref{sec:instantiation}, we define the operations used to instantiate a !-tensor into a concrete tensor and show that these operations preserve well-formedness. In Section~\ref{sec:box-reasoning}, we define !-tensor equations, which represent a families of equations between tensors, and provide inference rules to support rewriting. 
In Section~\ref{sec:examples}, we give some examples to show this !-box reasoning in action.

This is an expanded version of the conference paper~\cite{NoncommBB}. It has been updated to include a precise statement of soundness for the !-tensor inference rules, relevant proofs, and an additional example that illustrates some of the unique features of non-commutative !-box induction.


\section{Tensors}\label{sec:tensors}

Let $\mathcal C$ be a compact closed (a.k.a. symmetric monoidal autonomous) category, freely generated by a set of objects $X, Y, Z, \ldots$ and morphisms of the form $\phi : I \to X_1 \otimes \ldots \otimes X_n$, i.e. morphisms with only (non-trivial) outputs. $\mathcal C$ is compact closed, so this yields no loss of generality, as we can represent an input of type $A$ as an output of type $A^*$. For simplicity, we will ignore edge typing and assume every `input' is of fixed type $X^*$ and every `output' is of type $X$.

Since we want to distinguish individual inputs/outputs, we label them. We will use lower case letters to name edges. They will have a hat to illustrate being an `output': $\{\eout{a},\eout{b},\ldots\}$, or a check to illustrate being an `input': $\{\ein{a},\ein{b},\ldots\}$. Strings of these named and directed edges are written as subscripts on generating morphisms to fully describe them. For example, translating a morphism $\phi$ into tensor notation (with arbitrarily chosen edge names) yields:
\[
\phi : I \to X \otimes X \otimes X^* \otimes X^* \otimes X^* \qquad \leadsto \qquad
\term{\phi.+a+b-c-d-e.}
\]

We introduce a special graphical notation for morphisms with only outputs. We write them as circles with a tick, taking the convention that inputs/outputs are ordered clockwise from the tick.
\begin{equation}
  \term{\phi.+a+b-c-d-e.}
\quad := \quad
\input{phi-point.tikz}
\quad \leadsto \quad
\input{compl-fixed-arity.tikz}
\end{equation}

Writing two tensors side-by-side yields a new tensor formed by taking the monoidal product and \textit{contracting} (i.e.~plugging together) any repeated names using the compact structure on $X$. Hence in the following diagram $\eout{a}$ is plugged in to $\ein{a}$ and $\eout{b}$ is plugged in to $\ein{b}$.
\begin{equation}\label{eq:contraction}
  \term{\psi.+f\bR-a-b\e.\phi.\bR+a+b\e-c-d-e.}
\quad := \quad
\input{psi-phi-contract.tikz}
\quad \leadsto \quad
\input{compose-graphs.tikz}
\end{equation}

We say repeated edge names (e.g. $a$ and $b$ above) are \textit{bound} in a tensor expression, and all other edge names are \textit{free}. In the graph we have labelled the bound edges, though this is purely for demonstrating which edges are bound. The names of bound edges can be changed at will, provided they are replaced with new, fresh names. Hence $\term{\psi.+f\bR-a-b\e.\phi.\bR+a+b\e-c-d-e.}$ and $\term{\psi.+f\bR-x-y\e.\phi.\bR+x+y\e-c-d-e.}$ represent the same tensor (a notion which we explain in Definition~\ref{def:tensor-equiv}). As a result, we typically will not write down bound names in the graphical notation.

\begin{definition}
  The set of \textit{tensor expressions} for a signature $\Sigma$ consists of 
    \begin{align*}
      \bullet\; &\term{1} && \text{(empty tensor)} \\
      \bullet\; &\term{1.+a-b.} && \text{(identity tensor)} \\
      \bullet\; &\term{\psi.+a-b\ldots.} && \text{(generators from $\Sigma$)} \\
      \bullet\; &\term{GH} && \text{(for $G,H$ tensor expressions)}
    \end{align*}
  subject to the conditions that the arrangement of input/output names $\term{\psi.+a-b\ldots.}$ match the arity of $\psi$ in $\Sigma$, and $\eout{a}$ and $\ein{a}$ occur at most once for each name $a$.
\end{definition}

\begin{definition}\label{def:tensor-equiv}
	Two tensor expressions $G, G'$ are equivalent, written $G \equiv G'$ if $G$ can be made into $G'$ by replacing bound names or by applying one or more of the following identities:
  \[
    (GH)K \equiv G(HK) \qquad
    GH \equiv HG \qquad
    G1 \equiv G
  \]
  \[
    G\term{1.+b-a.} \equiv G[\ein b \mapsto \ein a] \qquad
    H\term{1.+a-b.} \equiv H[\eout b \mapsto \eout a]
	\]
  The first three are associativity, commutativity and unit rules for the product. Assume for the last two identities that $\ein b$ and $\eout b$ are free in $G$ and $H$, respectively. These two equivalences demonstate that plugging an edge in to an identity morphism is the same as renaming the edge. An $\equiv$-equivalence class of tensor expressions is called a \textit{tensor}.
\end{definition}

\begin{remark}
  We allow equivalent tensor expressions to be substituted for each other in the sense that if $G\equiv H$ then also $GK\equiv HK$ (assuming both expressions are well-formed). 
\end{remark}

Note that we use $\equiv$ for syntactic equivalence of tensor expressions (and later !-tensor expressions). We reserve the normal equals sign for equality by the rules of a given diagrammatic theory. As such, we always assume $(G \equiv H) \implies (G = H)$, i.e.~reflexivity modulo $\equiv$, but not the converse.

Tensors are related to morphisms in the free compact closed category as follows. Suppose we fix a set of \textit{canonical names} $\{ \ein{x}_1, \ein{x}_2, \ldots \}$ and $\{ \eout{x}_1, \eout{x}_2, \ldots \}$. A tensor $G$ is said to be \textit{canonically named} if for some $N$ it has as a free name precisely one of $\ein{x}_i$ or $\eout{x}_i$ for $1 \leq i \leq N$.

\begin{theorem}
	Canonically-named tensors are in 1-to-1 correspondence with points (i.e. morphisms $I \to ...$) in the free compact closed category $\mathcal C$.
\end{theorem}

\begin{proof}
  To prove this, we describe the construction depicted in~\eqref{eq:contraction} in more detail. This technique is very similar to the one employed in~\cite{KissingerATS}, but simpler in the compact closed case. First, we interpret a tensor expression as a morphism in the free compact closed category. We take the latter to be the category whose objects are lists in $X, X^*$ and whose morphisms are expressions in $\circ$, $\otimes$, symmetries $\sigma$ for $X, X^*$, identities $1_X, 1_{X^*}$, caps $\epsilon_X$, cups $\eta_X$, and the generators in $\Sigma$, modulo the (strict monoidal) compact closed axioms.

  Starting with a tensor expression, whose (canonical) free names are $x_1, \ldots, x_n$, we can choose bound names $a_1, \ldots, a_k$. Let $\psi^1, \ldots, \psi^m$ be either identities or generating morphisms. Then, we can interpret a tensor expression as:
  \begin{equation*}
    \psi^1_{...} \psi^2_{...} \ldots \psi^m_{...}
      \qquad\leadsto\qquad
     (1_* \otimes 1_* \otimes \ldots \otimes 1_* \otimes \epsilon_X \otimes \epsilon_X \otimes \ldots \otimes \epsilon_X) \circ
     \sigma \circ
     (\psi^1 \otimes \psi^2 \otimes \ldots \otimes \psi^m)
  \end{equation*}
  or graphically:
  \begin{equation}\label{eq:interp-form}
    \input{free-strict-form.tikz}
  \end{equation}
  where:
  \begin{enumerate}
    \item the $i$-th identity map is $1_X$ if the expression contains a free output $\hat{x_i}$ and $1_{X^*}$ if it contains a free input $\check{x_i}$,
    \item the $i$-th cap corresponds to the bound name $a_i$,
    \item $\psi^i$ is either $\eta_X$ for an identity tensor or the appropriate generator from $\Sigma$, and
    \item $\sigma$ is the (unique) map consisting just of symmetries and identities that connects the inputs/outputs of the $\psi^i$ associated with a given name to the appropriate identity or cap.
  \end{enumerate}
  The empty tensor expression is interpreted simply as $1_I$. This shows that a tensor expression uniquely determines a point in the free category. Since a tensor is an equivalence class of tensor expressions, we need to show that this doesn't depend on the choice of expression. We can safely ignore bracketing and instances of the empty tensor $1$, so if $G \equiv H$ it could be the case that $H$ differs from $G$ in (i) the order of the tensor symbols, (ii) the choice of bound names, or (iii) the number of identity tensors. In each of these cases, we use the axioms of a compact category to show that interpretations of $G$ and $H$ are equal.

  For (i), we can use naturality of $\sigma$ to reshuffle the generators at the bottom, without affecting connectivity. For (ii), we do the same thing, but for the caps at the top. For (iii), we can use the compactness equations:
  \ctikzfig{compactness}
  to insert or remove identities, i.e. cups, from the bottom.

  Conversely, any expression in the free category can be written in the form of~\eqref{eq:interp-form}, at which point one can read off the tensor expression. First, use bifunctoriality to pull all generators and cups to the bottom and all caps to the top. Then, use naturality to sort all of the caps to the right of the output wires. The only freedom is in the order of the generators/cups and the caps, which is captured by $\equiv$.
\end{proof}

As mentioned at the start of this section, it suffices to consider points, as non-trivial inputs of type $X$ can be turned, via the compact structure, into outputs of type $X^*$. To summarise, we can interpret a tensor in a compact closed category as follows. First, we swap its free names for `canonical names' (or otherwise order the outputs somehow), then interpret each atomic expression as a morphism. Finally, we construct the composed morphism by composing each of the components and contracting repeated edge names, as in~\eqref{eq:contraction}.

Alternatively, one can study models in an existing abstract tensor system (in the sense of Penrose), in which case interpretation is trivial. These two points of view (categorical vs. ATS) are equivalent, up to the use of order vs. names to distinguish free edges, as was shown in~\cite{KissingerATS}.

\section{Adding !-boxes to tensor expressions}\label{sec:bboxes}

We now extend the existing tensor notation with !-boxes. Graphically !-boxes are blue boxes surrounding a subgraph, labelled with a name ($A,B,\ldots$). We can denote this with square brackets around a subexpression in a tensor expression, labelled with the !-box's name. Intuitively a !-box represents a portion of the graph that can be copied multiple times. For this to be well-defined in the non-commutative case we need to clarify where each new copy of the subexpression gets attached to surrounding nodes.

This is done by assigning an expansion direction (clockwise vs anticlockwise) to any group of edges from a node to a !-box. We denote clockwise edge groups in tensor expressions as $\eclock{\dots}{A}$ and anticlockwise edge groups as $\eanti{\ldots}{A}$. Graphically, we depict this as directed arcs over groups of edges. For example:
\[
\term{\phi.<+a]B.[\psi.-a.]B} \ :=\  \input{bb-example.tikz} \qquad\textrm{vs.}\qquad
\term{\phi.[+a>B.[\psi.-a.]B} \ :=\  \input{bb-example-cw.tikz}
\]
The arc notation clarifies not only which direction edges should expand, but also whether they should expand in groups or individually. For example, the following diagrams demonstrate notation for anti-clockwise expansion of $\term{+a-b}$ as a group, clockwise expansion of $\term{+a-b}$ as a group, and clockwise expansion of $\term{+a}$ and $\term{-b}$ as individual edges, respectively: 
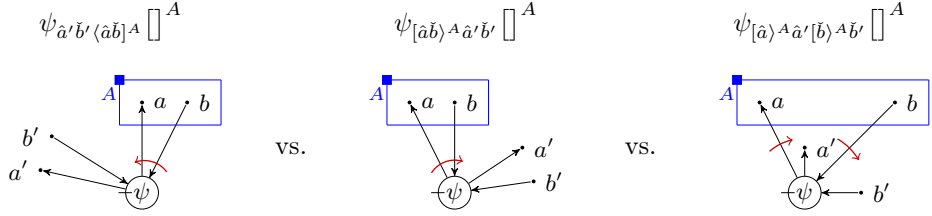
\[
\begin{matrix}
  \term{\psi.+a'-b'<+a-b]A.[]A} & &
  \term{\psi.[+a-b>A+a'-b'.[]A} & &
  \term{\psi.[+a>A +a'[-b>A -b'.[]A} \\
  & & & & & \\
  \input{expand-order1.tikz} &
  \quad\textrm{vs.}\quad &
  \input{expand-order2.tikz} &
  \quad\textrm{vs.}\quad &
  \input{expand-order3.tikz}
\end{matrix}
\]

It is also possible for !-boxes to be nested inside of other !-boxes. This means expansion of the parent !-box makes a new copy of the child with a new !-box name. Edge groups can correspondingly be nested if the edges enter more than one !-box. In the diagram below we have the !-graph corresponding to a !-tensor with nested !-boxes.
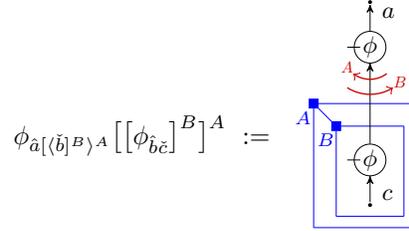
\[ \term{\phi.+a[<-b]B>A.[[\phi.+b-c.]B]A} \ :=\ 
\input{nested-ex.tikz} \]
Note that we have labelled each arc with its associated !-box. This is not necessary if we adopt the convention that arcs corresponding to !-boxes are always drawn \textit{closer} to the node than their children.



\begin{definition}\label{def:edgeterm-equiv}
Fix disjoint, infinite sets $\edgenames$ and $\boxnames$ of edge names and !-box names, respectively. We denote the set of \textit{directed edges} as $\diredgenames := \{\ein{a},\eout{a} : a\in\edgenames\}$. The set of \textit{edgeterms} $\edgeterms$ is defined recursively as follows:
	\begin{align*}
		\bullet\; &\epsilon \in \edgeterms && \text{(empty edgeterm)} \\
		\bullet\; & \ein a, \eout a \in \edgeterms && a \in \edgenames \\
		\bullet\; &\term{[e>A},\term{<e]A} \in \edgeterms && e\in\edgeterms,\; A\in\boxnames \\
		\bullet\; & e f \in \edgeterms && e,f\in\edgeterms
	\end{align*}
	Two edgeterms are equivalent if one can be transformed into the other by:
	\begin{equation}\label{eqn:edgeterm_equiv}
    e(fg) \equiv (ef)g \qquad
    \epsilon e \equiv e \equiv e \epsilon \qquad
    \term{[\epsilon>A} \equiv \epsilon \equiv \term{<\epsilon]A}
  \end{equation}
  The first two represent associativity and identity of concatenation, with $\epsilon$ as the unit. The last equivalence says that empty edge groups can be ignored.
\end{definition}

Since the well-formedness conditions for !-tensor expressions are a bit more complicated than for tensor expressions, we first define the set of all !-pretensor expressions, including those that may be ill-formed.

\begin{definition}
The set of all !-pretensor expressions $\graphterms'$ for a signature $\Sigma$ is defined recursively as:
  \begin{align*}
    \bullet\; &1, \term{1.+a-b.} \in \graphterms' && a,b\in\edgenames\\
    \bullet\; &\term{\phi.e.} \in \graphterms' && e\in\edgeterms, \phi\in\Sigma \\
    \bullet\; &\term{[G]A} \in \graphterms' && G\in\graphterms', \; A\in\boxnames \\
    \bullet\; & G H \in \graphterms' && G,H\in\graphterms'
  \end{align*}
\end{definition}

We introduce the notion of a \textit{context}, which lists the !-boxes in which a certain edge name occurs, from the inside-out. These come in two flavours, \textit{edge contexts} and \textit{node contexts}.

\begin{definition}
  Given a directed edge $a\in\diredgenames$ in a !-tensor $G$ nested as $\term{[[\phi.<<a>{E_1}\ldots>{E_n}.]{N_1}\ldots]{N_m}}$.

  We define the \textit{edge context}, \textit{node context}, and \textit{context} of $a$ respectively as:
    \begin{align*}
      \ectx_G(a)&:=[E_1,\ldots,E_n] && \text{(edge context)} \\
      \nctx_G(a)&:=[N_1,\ldots,N_m] && \text{(node context)} \\
      \ctx_G(a)&:=\ectx_G(a).\nctx_G(a) && \text{(context)}
    \end{align*}
    That is, $\ectx_G(a)$ lists the !-boxes containing $a$ that occur as part of $a$'s edgeterm, and $\nctx_G(a)$ lists the rest. 
\end{definition}

Finally, a !-tensor expression is a !-pretensor expression where !-box/edge names must be suitably unique and occur in compatible contexts:

\begin{definition}\label{def:tensor_conditions}
  A !-tensor expression is a !-pretensor expression satisfying the following conditions:
  \begin{itemize}
    \item[F1.] $\ein a$ and $\eout a$ occur at most once for each edge name $a$
    \item[F2.] $\term{[\ldots]A}$ must occur at most once for each !-box name $A$
    \item[C1.] $\ectx_G(a)\cap \nctx_G(a)=\varnothing$ \: for all edges $a \in \edgenames$ in $G$
    \item[C2.] If $\ectx_G(a)=[B_1,\ldots,B_n]$ \; then all $B_i\in\Boxes(G)\:$ and $\; B_1\prec_G B_2\prec_G \ldots \prec_G B_n$
    \item[C3.] For all bound pairs $\ein a, \eout a$ of edge names in $G$, there exist lists $es, bs$ of !-box names such that:
    \[ es.\nctx_G(\ein{a}) = \ectx_G(\eout{a}).bs
       \quad \textrm{and} \quad
       es.\nctx_G(\eout{a}) = \ectx_G(\ein{a}).bs \]
  \end{itemize}
  where $A \prec_G B$ means that the !-box $A$ is nested immediately inside $B$ in $G$ (i.e.~without other !-boxes nested between). We write $\graphterms$ for the set of all !-tensor expressions for a signature $\Sigma$.
\end{definition}

The freshness conditions F1 and F2 ensure that we have not used the same name for more than one (directed) edge and that each !-box name appears only once. If a node is in !-box $B$ then any edges attached to it are already in $B$ so it wouldn't make sense to have $B$ in both the $\ectx(a)$ and $\nctx(a)$ for $a\in\diredgenames$, this is enforced by C1. C2 ensures that edge contexts are compatible with the !-boxes in the rest of the !-tensor expression. For example $\term{\phi.[[-a>A>B.}$ requires $A$ to be nested in $B$ so does not result in a valid expression when composed with e.g. $\term{[[\psi.-b.]B]A}$. 
C3 ensures that edges into !-boxes from the outside are decorated correctly by their edge terms. For instance, this is allowed: $\term{\psi.[+a>A.[\phi.-a.]A}$ but this is not: $\term{\psi.+a.[\phi.-a.]A}$. The freedom to pick $bs, es$ allows bound pairs of edges to share some common context, e.g.: $\term{[\psi.+a.\phi.-a.]A}$ (both nodes, and hence the edge, are inside $A$) or $\term{\psi.[+a>A.\phi.<-a]A.[]A}$ (only the edge is inside $A$). In the second example, $A$ occurs in an edge term, so C2 requires the presence of $\term{[\ldots]A}$ somewhere in the !-tensor expression, hence we append the empty !-box $\term{[]A}$ (actually shorthand for $\term{[1]A}$). In particular, empty !-boxes are meaningful, which justifies our decision not to include the equivalence $\term{[1]A}\equiv 1$ in Definition~\ref{def:tensor-equiv}.


In this paper when we write a composition $GH$, unless stated otherwise, we will assume this forms a well defined !-tensor expression.

Naturally, we can say two !-tensor expressions are equivalent, written $G \equiv H$, if one can be obtained from the other by using the usual tensor equivalences from Definition~\ref{def:tensor-equiv} and/or edgeterm equivalences from \eqref{eqn:edgeterm_equiv}. However, we need to generalise the last two tensor equivalences to allow $\term{1.+b-a.},\term{1.+a-b.}$ to appear inside !-boxes: 
\begin{align}
\label{eqn:contract_equiv}
\begin{split}
  \term{G[K_n...[K_1 1.+b-a.]{B_1}...]{B_n} \equiv G[-b \mapsto -a] [K_n...[K_1]{B_1}...]{B_n}} \\
  \term{H[K_n...[K_1 1.+a-b.]{B_1}...]{B_n} \equiv H[+b \mapsto +a] [K_n...[K_1]{B_1}...]{B_n}}
\end{split}
\end{align}
where $\ein b$ and $\eout b$ are free in $G$ and $H$ respectively. These allow identities connected to nodes outside of !-boxes to still be simplified. For example:
\[ \term{\psi.[+a>A.[1.+b-a.]A} \equiv \term{\psi.[+b>A.[]A} \]


When we talk about !-tensors we are not interested in the explicit !-tensor expression, only the equivalence class of such expressions under $\equiv$. It is these classes that correspond to diagrams in our !-box graphical notation.

\begin{remark}
  We allow equivalent !-tensor expressions to be substituted for each other in the sense that $G\equiv H$ implies $GK\equiv HK$ (as with tensor expressions) and $\term{[G]A}\equiv\term{[H]A}$. Thus, $G\equiv H$ implies, for example, that $\term{[K_n\ldots [K_1 G]{B_1}\ldots]{B_n}}\equiv\term{[K_n\ldots [K_1 H]{B_1}\ldots]{B_n}}$.
\end{remark}

We call the graphical notation for !-tensors the \textit{non-commutative !-graph notation}, or simply !-graphs.

\begin{theorem}\label{thm:bang-graph-notation}
  Any !-tensor can be represented unambiguously using non-commutative !-graph notation.
\end{theorem}

\begin{proof}
  We show this by providing a general procedure for interpreting a !-graph as a !-tensor expression, and vice-versa. For the sake of clarity, we demonstrate each step on a worked example. Given a non-commutative !-graph, we wish to obtain a unique equivalence class of !-tensor expressions under $\equiv$. Begin by choosing fresh names to write on all the interior edges.
  \[ \input{bb-interp1.tikz} \quad \Longrightarrow \quad
     \input{bb-interp2.tikz} \]
  Then, write the !-boxes with nesting as depicted in the diagram:
  \[ \term{\ldots[\ldots]C[\ldots[\ldots]B]A} \]
  Write each node in the diagram on the location it occurs (w.r.t. !-boxes):
  \[ \term{\phi.\ldots.[\psi.\ldots.]C[[\psi.\ldots.]B]A} \]
  Finally, add the edges of each node, reading clockwise from the tick. Edges occurring under a clockwise arrow marked $A$ should be enclosed in $\eclock{\dots}{A}$, and edges under an anti-clockwise arrow should be enclosed in $\eanti{\dots}{A}$, where the outermost groups are the ones closest to the node in the picture.
  \[ \term{\phi.+a[<-e]B>A<-d]C.[\psi.+d-c.]C[[\psi.+e-b.]B]A} \]
  The only choices we made in this process were the choice of interior edge names and the order in which to write the individual tensors. However, up to $\equiv$, these are irrelevant. To show that any !-tensor can be represented this way, we simply run the above procedure in reverse.
\end{proof}

Because of this theorem, we use the terms !-tensor and !-graph interchangeably, depending on whether we wish to refer to the syntactic vs. graphical notation. 

\section{Instantiating tensor expressions with !-boxes}\label{sec:instantiation}

The following diagram demonstrates two !-box operations we can apply to a graph: killing a !-box is the operation deleting the !-box $B$ and all contents (including edges to/from $B$), and expanding is the operation creating a new concrete instance of the subgraph inside $B$ (attached appropriately). Below each diagram is a possible !-tensor expression representing it.

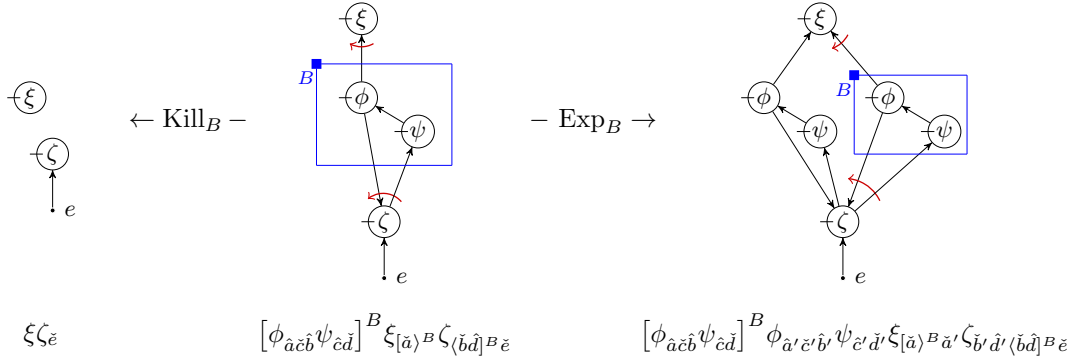
\begin{center}
  \input{bb-ex1-kill.tikz}
  \quad$\leftarrow \Kill_B -$
  \input{bb-ex1.tikz}
  $-\, \Exp_B \rightarrow$\hspace{-10pt}
  \input{bb-ex1-exp.tikz}
\end{center}

From a !-tensor we can use these two operations to create any number of copies of the contents of a !-box. There are, however, other operations which allow us to prove more powerful !-tensor equations. Two other such operations we shall use are $\Copy_B$ and $\Drop_B$. $\Copy_B$ makes a new copy of the !-box $B$, including the !-box itself with a new fresh name. $\Drop_B$ removes the !-box $B$ but leaves it's contents as they were. Some of these operations involve copying various edge/!-box names, so we need a means of obtaining fresh names if we wish to give formal definitions.

\begin{definition}
  Given names $E\subset\edgenames,B\subset\boxnames$, a \textit{freshness function} for the pair $(E,B)$ is a pair of bijections $\fr: \edgenames \to \edgenames$ and $\fr: \boxnames \to \boxnames$ such that
  \[ E \cap \fr(E) = \varnothing \quad \textrm{and} \quad
     B \cap \fr(B) = \varnothing \]
\end{definition}

Freshness functions provide unique new names for sets of edge and !-box names, usually taken from some given !-tensor expressions. Since (finitely many) !-tensor expressions only need finitely many names and $\edgenames$ and $\boxnames$ are both infinite, we never need to worry about existence of such a function. We will often state that $\fr$ is `an appropriate freshness function'. Let $\Edges(G) \subset \edgenames$ and $\Boxes(G) \subset \boxnames$ be the edge names and !-box names occurring in a !-tensor $G$, respectively. Any freshness function for $(\Edges(G),\Boxes(G))$ is then an appropriate freshness function for expanding a !-box in $G$.

For !-tensor expressions $G$ or edgeterms $e$, we will write $\fr(G)$ or $\fr(e)$ to designate the new expressions with names substituted according to the given bijection.

\begin{definition}
  A !-box operation $\Op_B$ (acting on !-box $B$) is an operation on !-tensor expressions which acts somewhat trivially on the following cases:
  \begin{align*}
    \Op_B(G H) & := \Op_B(G) \Op_B(H) &
    \Op_B(e f) &:= \Op_B(e) \Op_B(f) \\
    \Op_B(\term{[G]A}) &:= \term{[\Op_B(G)]A} &
    \Op_B(\term{[e>A}) &:= \term{[\Op_B(e)>A} \\
    \Op_B(\phi_e) &:= \phi_{\Op_B(e)} &
    \Op_B(\term{<e]A}) &:= \term{<\Op_B(e)]A} \\
    \Op_B(x) & : = x & &
  \end{align*}
  where $A \neq B$ and $x \in \{ 1, \term{1.+a-b.}, \ein a, \eout a, \epsilon \}$.
  The !-box operations we are interested in are the following (defined by their actions in the three cases not mentioned above):
  \begin{align*}
    \Exp_B(\term{[G]B})  &:= \term{[G]B\fr(G)} &
    \Kill_B(\term{[G]B}) &:= 1 \\
    \Exp_B(\term{[e>B})  &:= \term{[e>B\fr(e)} &
    \Kill_B(\term{[e>B}) &:= \epsilon \\
    \Exp_B(\term{<e]B})  &:= \term{\fr(e)<e]B} &
    \Kill_B(\term{<e]B}) &:= \epsilon
  \end{align*}
  \begin{align*}
    \Copy_B(\term{[G]B})  &:= \term{[G]B[\fr(G)]{\fr(B)}} &
    \Drop_B(\term{[G]B})  &:= \term{G} \\
    \Copy_B(\term{[e>B})  &:= \term{[e>B[\fr(e)>{\fr(B)}} &
    \Drop_B(\term{[e>B})  &:= \term{e} \\
    \Copy_B(\term{<e]B})  &:= \term{<\fr(e)]{\fr(B)}<e]B} &
    \Drop_B(\term{<e]B})  &:= \term{e}
  \end{align*}
\end{definition}

Note that $\Exp_B$ and $\Copy_B$ implicitly take freshness functions as input. If we wish to make this explicit, we will write $\Exp_{B,\fr}$ or $\Copy_{B,\fr}$ respectively. Note that our four operations are not minimal in that some can be written in terms of others. For example it is easy to see that $\Exp_{B,\fr}=\Drop_{\fr(B)}\circ\Copy_{B,\fr}$.

The above operations can be lifted from !-tensor expressions to !-tensors, i.e. $\equiv$-classes of expressions, because of Theorem~\ref{thm:operation_equivalence}.

\begin{theorem}\label{thm:ops_create_tensors}
  If $G\in\graphterms$ and $\Op_A$ is one of our !-box operations then $\Op_A(G)\in\graphterms$. 
\end{theorem}
\begin{proof}
  (Sketch) Shown by checking that the conditions F1-2, C1-3 still hold for each operation. F1-2 are trivial since new edges/!-boxes have new names. C1-3 are checked for each defined !-box equation in Appendix~\ref{app:ops_create_tensors}.
\end{proof}

\begin{theorem}\label{thm:operation_equivalence}
  Let $\fr$ be a freshness function for the !-tensor expressions $G, H$. Then $G\equiv H$ implies $\Op_{B,\fr}(G) \equiv \Op_{B,\fr}(H)$.
\end{theorem}
\begin{proof}
  We need to check our enforced equivalences still hold after $\Op_B$. It is clear from the definitions of $\Op_B(G H), \Op_B(e f), \Op_B(1), \Op_B(\epsilon)$ that associativity/commutativity/unit conditions are preserved. The others need to be checked individually. We present the proof in Appendix~\ref{app:ops_pres_equiv}.
\end{proof}

\begin{definition}\label{def:!tensor_semantics}
  A tensor $G'$ is a \textit{concrete instance} of a !-tensor $G$ if it is obtained from $G$ by repeatedly applying the two !-box operations $\Exp$ and $\Kill$ until there are no !-boxes left. This sequence of operations is called the \textit{instantiation} of $G'$. We write $\inst{G}$ for the set of instantiations of $G$ and $\sem{G}:=\{i(G):i\in\inst{G}\}$ for the set of all concrete instances.
\end{definition}

\begin{theorem}\label{thm:top_op_first}
  Given an instantiation $i\in\inst{G}$ and a top-level !-box $A$ (i.e with no parent !-box) in $G$, $i$ can be rewritten as $i'\circ\Kill_A\circ\Exp_A^n$ where $i'\in\inst{\Kill_A\circ\Exp_A^n(G)}$.
\end{theorem}
\begin{proof}
  We need to check that operations on $A$ can always be commuted to the right, past other operations. If $B$ is not nested in $A$, it is easy to show that $\Op_A\circ\Op_B = \Op_B\circ\Op_A$. Otherwise, supposing $B$ is nested in $A$:
  \begin{itemize}
    \item If $\Op_A=\Kill_A$ then killing $A$ will erase any part of the graph resulting from $\Op_B$, i.e. $\Kill_A \circ \Op_B = \Kill_A$.
    \item Otherwise $\Op_A = \Exp_{A,\fr}$, in which case: $\Exp_{A,\fr} \circ \Op_B = \Op_{\fr(B)} \circ \Op_B \circ \Exp_{A,\fr}$. In the case that $\Op_B = \Exp_B$, freshness functions on the RHS need to be chosen to produce identical names to the LHS.
  \end{itemize}
\end{proof}

\begin{notation}
  We will write $\KE^n_A$ as a shorthand for $\Kill_A\circ\Exp_A^n$.
\end{notation}

\begin{corollary}
  Given a total order on !-box names, !-graph instantiations admit a normal form.
\end{corollary}

\begin{proof}
  Given an instantiation of $G$ we can chose the first (ordered by !-box name) top-level !-box, $A$, and move it's operations (of the form $\KE_A^n$) completely to the right. The rest is then an instantiation of $\KE_A^n(G)$, so we can repeat the process. Termination of this procedure can be shown since each step removes a !-box and only adds !-boxes with fewer levels of nesting.
\end{proof}

When we fix a model in some category $\mathcal C$, concrete tensors can then be interpreted as morphisms in $\mathcal C$, just as before. Unlike before, we allow the generators in $\Sigma$ to have variable arities. We interpret the variable-arity generators as sets of morphisms, indexed (uniquely) by their configuration of input/output edges. The arity of a generator is a word over $\{\wedge,\vee\}$ where $\wedge$ represents an output and $\vee$ an input. For example the node $\term{\psi.+a-b+c+d.}$ 
has an arity $\wedge\vee\wedge\:\wedge$ and needs to be assigned a morphism $f : I\to X \otimes X^* \otimes X \otimes X$. Thus, a single generator is interpreted as a function of the type $\phi:\{\wedge,\vee\}^*\to \text{Mor}(\mathcal{C})$. Extending this to !-tensors, a single !-tensor can be interpreted in $\mathcal C$ as a set of morphisms, one for each concrete instance.

\section{Reasoning with !-boxes}\label{sec:box-reasoning}

The real power of !-boxes comes from the ability to do equational reasoning using infinite families of rules. Just as it makes sense to instantiate a single !-tensor, it makes sense to instantiate an \textit{equation} $G = H$ between two !-tensors, provided they have compatible boundaries.

\begin{definition}
  A \textit{!-tensor equation} `$G = H$' consists of a pair of !-tensors $(G, H)$ that have \textit{compatible boundaries}. That is, they have identical free edge names and !-boxes, $A \prec_G B \Leftrightarrow A \prec_H B$ for all !-boxes in $G$ and $H$, and $\ctx_G(a) = \ctx_H(a)$ for all free edge names.
\end{definition}

Intuitively, we require that the LHS and RHS of a !-tensor equation have the same interface to attach to other graphs (same free variables and same !-box structure). These consistency conditions guarantee that (i) applying !-box operations to valid equations yields valid equations, and (ii) when $G$ occurs as a sub-expression of some other !-tensor $K$, it can be substituted for $H$ to yield another valid !-tensor $K'$.

\begin{theorem}
  Let $\fr$ be a appropriate freshness function for !-tensors $G, H$. Then, if $G = H$ is a !-tensor equation, then so too is:
  \[{\Op_B(G = H)\ \ :=\ \  (\Op_B(G) = \Op_B(H))}\]
  Where $\Op_B$ is any of our four !-box operations and if required we have used the freshness function $\fr$.
\end{theorem}
\begin{proof}
  From the table in Lemma~\ref{lem:contexts}, we see that $\ctx_{\Op_B(G)}(a)$ and (if defined) $\ctx_{\Op_B(G)}(\fr(a))$ depend only on the original context $\ctx_G(a)$. Thus, if $\ctx_G(a)=\ctx_H(a)$, then $\ctx_{\Op_B(G)}(a)=\ctx_{\Op_B(H)}(a)$ and (if defined) $\ctx_{\Op_B(G)}(\fr(a))=\ctx_{\Op_B(H)}(\fr(a))$.
\end{proof}


We let $\inst{G=H} := \inst{G}$, which is the same by compatibility as letting $\inst{G=H} := \inst{H}$. As in the case of !-tensors, we can then define $\llbracket G = H \rrbracket$ to be the set of all concrete rules derivable from $G = H$ using the !-box operations, i.e. $\llbracket G = H \rrbracket := \{i(G)=i(H):i\in\inst{G=H}\}$. A valid model of a graphical theory $(\Sigma, E)$ is one where all of the equations in $\llbracket G = H \rrbracket$ hold for each equation $G = H$ in $E$. Proving that a rule holds for \textit{all} of its instances could be a daunting task in general, however in many cases a technique called \textit{!-box induction}---which we will meet shortly---comes to the rescue.

We obtain a notion of substitution of sub-expressions constructively, via inference rules. We assume that $=$ contains $\equiv$, and is furthermore symmetric and transitive. We additionally assume that it lifts over products and !-boxes:
\begin{equation}
    \scalebox{1.0}{
  \AxiomC{$G = H$}
  \RightLabel{\scriptsize(Prod)}
  \UnaryInfC{$GK = HK$}
  \DisplayProof
\qquad\qquad
  \AxiomC{$G=H$}
  \RightLabel{\scriptsize(Box)}
  \UnaryInfC{$\term{[G]A} = \term{[H]A}$}
  \DisplayProof
  }
\end{equation}
where we assume that $K$ and $A$ are chosen such that $GK$, $HK$, $\term{[G]A}$, and $\term{[H]A}$ are well-defined.
Since $GK = KG, HK = KH$, the product rule could equivalently be written with the substitution on the right, or on both sides:
\begin{prooftree}
  \AxiomC{$G=G'$}
  \AxiomC{$H=H'$}
  \RightLabel{\rm\scriptsize(Prod')}
  \BinaryInfC{$GH=G'H'$}
\end{prooftree}




These rules provide the conditions under which some equation $G = H$ can be unified, given some context, with a bigger equation $G' = H'$. The final inference rule (Weaken) is less intuitive from the point of view of terms, and is best understood graphically. Consider the following embedding of !-graphs:
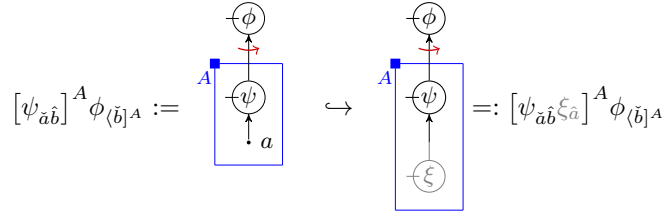
\[ \term{[\psi.-a+b.]A\phi.<-b]A.} :=
   \input{embed-ex.tikz} =:
   \term{[\psi.-a+b.\color{black!50!white}\xi.+a.\color{black}]A{\phi.<-b]A.}} \]
The !-tensor on the left does not embed as a sub-term of the one on the right, because the !-box $A$ has more nodes on the right. However, semantically, this is perfectly fine, as all of the concrete instances of the left !-tensor will have (uniquely-determined) embeddings into all of the concrete instances of the right one. So, we also need a rule that allows us to `weaken' !-boxes by adding more nodes to them.
\begin{equation}\label{eq:weaken-inf}
  \scalebox{1.0}{
  \AxiomC{$G=G'$}
  \RightLabel{\scriptsize(Weaken)}
  \UnaryInfC{$\Wk{A}{K}(G)=\Wk{A}{K}(G')$}
  \DisplayProof
  }
\end{equation}
where $\Wk{A}{K}(G)$ is defined recursively as:
\begin{align*}
  \Wk{A}{K}(\term{[G]A}) &:= \term{[GK]A} \\
  \Wk{A}{K}(\term{[G]B}) &:= \term{[\Wk{A}{K}(G)]B} & \text{if } A\not=B \\
  \Wk{A}{K}(G H) &:= \Wk{A}{K}(G) \Wk{A}{K}(H) \\
  \Wk{A}{K}(x) &:= x & x \in \{ 1, \term{1.+a-b.}, \phi_e \}
\end{align*}

Just like with products, when we write $\Wk{A}{K}(G)$ we assume that $K$ is chosen such that this is a well-formed !-tensor (unless stated otherwise).
We can now show that weakening respects $\equiv$, so it lifts to an operation on $\equiv$-classes of !-tensor expressions:

\begin{theorem}\label{thm:weaken_tensors}
  $G\equiv H$ implies $\Wk{A}{K}(G)\equiv\Wk{A}{K}(H)$.
\end{theorem}
\begin{proof}
  We need to check our enforced equivalences still hold after $\Wk{A}{K}$. It is clear from the definitions of $\Wk{A}{K}(G H), \Wk{A}{K}(1)$ that associativity/commutativity/unit conditions are preserved. We check the other two cases:
  \begin{itemize}
    \item We need to check the equivalence $\term{G[K_n...[K_1 1.+b-a.]{B_1}...]{B_n} \equiv G[-b \mapsto -a] [K_n...[K_1]{B_1}...]{B_n}}$ (with $\term{-b}$ free in $G$) is preserved by weakening. \\
    If $A\not\in[B_1,\ldots,B_n]$ then neither $\term{+b}$ nor $\term{-b}$ is affected by $\Wk{A}{K}$ so we get:
    \begin{align*}
      \term{\Wk{A}{K}(G[K_n...[K_1 1.+b-a.]{B_1}...]{B_n})} &\equiv \term{\Wk{A}{K}(G)[\Wk{A}{K}(K_n)...[\Wk{A}{K}(K_1) 1.+b-a.]{B_1}...]{B_n})} \\
          &\equiv \term{\Wk{A}{K}(G)[-b \mapsto -a] [\Wk{A}{K}(K_n)...[\Wk{A}{K}(K_1)]{B_1}...]{B_n})} \\
          &\equiv \term{\Wk{A}{K}(G[-b \mapsto -a] )\Wk{A}{K}([K_n...[K_1]{B_1}...]{B_n}))} \\
          &\equiv \term{\Wk{A}{K}(G[-b \mapsto -a] [K_n...[K_1]{B_1}...]{B_n})}
    \end{align*}
    Otherwise $A=B_i$ for some $i\leq n$. In which case:
    \begin{align*}
      \term{\Wk{A}{K}(G[K_n...[K_1 1.+b-a.]{B_1}...]{B_n})}
      &\equiv \term{G[K_n...[K_i K ...[K_1 1.+b-a.]{B_1}...]{B_i}...]{B_n}} & \\
          &\equiv \term{G[-b \mapsto -a] [K_n...[K_i K ...[K_1]{B_1}...]{B_i}...]{B_n}} & \\
          &\equiv \term{\Wk{A}{K}(G[-b \mapsto -a] [K_n...[K_1]{B_1}...]{B_n})} &
    \end{align*}
    \item We need to check $\term{H[K_n...[K_1 1.+a-b.]{B_1}...]{B_n} \equiv H[+b \mapsto +a]}$ is preserved for $\term{+b}$ free in $G$. \\
    This proof is similar to that above.
  \end{itemize}
\end{proof}

For~\eqref{eq:weaken-inf} to make sense, we also need to check that weakening preserves compatibility:

\begin{theorem}\label{thm:weaken_compat}
  If $G$ and $H$ are compatible !-tensors then so are $\Wk{A}{K}(G)$ and $\Wk{A}{K}(H)$.
\end{theorem}
\begin{proof}
  Let $as$ be the context in which $K$ is added to the !-tensors $G,H$. So if $A\prec_G B \prec_G C$ then $as:=[A,B,C]$. Then for an edge $a$:
  \begin{align*}
    \ctx_{\Wk{A}{K}(G)}(a) &= \begin{cases}
                                \ctx_G(a) &a\not\in K \\
                                \ctx_K(a).as &a\in K
                              \end{cases} \\
                           &= \begin{cases}
                                \ctx_H(a) &a\not\in K \\
                                \ctx_K(a).as &a\in K
                              \end{cases} \\
                           &= \ctx_{\Wk{A}{K}(H)}(a)
  \end{align*}
  It is easy to see that the !-box structure of $\Wk{A}{K}(G)$ is equal to the !-box structure of $G$ but with the structure of $K$ added inside $A$. The same is true of $\Wk{A}{K}(H)$ and $H$, so !-box structure equality is preserved.
\end{proof}

Here is an example of using the (Weaken) rule:
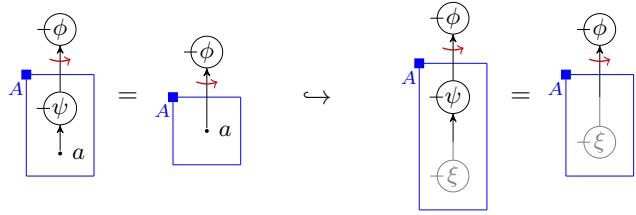
\[ \input{embed-rule-ex.tikz} \qquad \hookrightarrow \qquad
   \input{embed-big-rule-ex.tikz}  \]
The equation on the left is applied to delete all of the $\psi$-nodes occurring as input to a $\phi$, even though there may be more nodes in the !-box $A$.

A standard ingredient we need for unification is the ability to rename free edge and !-box names:
\begin{equation}\label{eq:renamers}
\AxiomC{$G = H$}
    \RightLabel{\rm\scriptsize(EdgeRename)}
    \UnaryInfC{$G[a\mapsto b] = H[a\mapsto b]$}
    \DisplayProof
\qquad\qquad
\AxiomC{$G = H$}
  \RightLabel{\rm\scriptsize(BoxRename)}
  \UnaryInfC{$G[A\mapsto B] = H[A\mapsto B]$}
  \DisplayProof
\end{equation}
It turns out that (EdgeRename) already follows from our existing rules, and with the addition of rules for $\Copy$ and $\Kill$, which we will introduce shortly, we can derive (BoxRename). See Appendix~\ref{sec:app-renaming} for details.

By iterating these rules, we can rename using any function $\rn:\edgenames\rightarrow\edgenames$ or $\rn:\boxnames\rightarrow\boxnames$ with finite support (i.e $\rn(x)=x$ for all but finitely many $x$).
We write $G[\rn]$ for $G$ with $\rn$ applied to all free edge (or !-box) names. Then, from~\eqref{eq:renamers}, it follows that:
\begin{equation}\label{eq:renamers2}
  \AxiomC{$G = H$}
  \RightLabel{\rm\scriptsize(EdgeRename')}
  \UnaryInfC{$G[\rn] = H[\rn]$}
  \DisplayProof
  \qquad\qquad
  \AxiomC{$G = H$}
  \RightLabel{\rm\scriptsize(BoxRename')}
  \UnaryInfC{$G[\rn] = H[\rn]$}
  \DisplayProof
\end{equation}

To allow (partial) instantiation of equations, we also include inference rules for each of our !-box operations i.e.
\[
  \AxiomC{$G = H$}
  \RightLabel{\scriptsize($\Exp_B$)}
  \UnaryInfC{$\Exp_B(G = H)$}
  \DisplayProof
\hspace{80pt}
  \AxiomC{$G=H$}
  \RightLabel{\scriptsize($\Kill_B$)}
  \UnaryInfC{$\Kill_B(G = H)$}
  \DisplayProof
\]
\[
  \AxiomC{$G=H$}
  \RightLabel{\scriptsize($\Copy_B$)}
  \UnaryInfC{$\Copy_B(G = H)$}
  \DisplayProof
\hspace{80pt}
  \AxiomC{$G=H$}
  \RightLabel{\scriptsize($\Drop_B$)}
  \UnaryInfC{$\Drop_B(G = H)$}
  \DisplayProof
\]
and an introduction rule for new !-boxes, called \textit{!-box induction}. Induction requires that $A$ is a top-level !-box (i.e. it has no parent !-box):
\[
	\scalebox{1.0}{
  \AxiomC{$\Kill_B(G=H)$}
  \AxiomC{$G=H\vdash_B\Exp_B(G=H)$}
  \RightLabel{\scriptsize($\textrm{Induction}_B$)}
  \BinaryInfC{$G=H$}
  \DisplayProof
  }
\]

Hence to prove $G=H$ we can prove a base case, $\Kill_B(G=H)$, and a step case which shows essentially that an equation involving $n$ copies of the contents of $B$ yields an equation with $n+1$ copies. 
The notation $G=H \vdash_B \Exp_B(G=H)$, means that $G = H$ can be used in the proof of $\Exp_B(G=H)$, provided that no !-box operations are applied to $B$.

\begin{remark}\label{rem:bang-logic}
  The utility of $\vdash_B$ is provided in \cite{MerryThesis} by an operation called \textit{fixing}, applied directly to !-graphs. In some sense, this is a stopgap until a proper logical system involving !-boxes is developed. This new `!-logic' is the topic of a forthcoming paper by the authors.
\end{remark}

\begin{theorem}
  The rules {\rm (Prod)}, {\rm (Box)}, {\rm (Weaken)}, $(\Exp_B)$, $(\Kill_B)$, $(\Copy_B)$, $(\Drop_B)$, {\rm (EdgeRename)}, and {\rm (BoxRename)} are sound with respect to $\llbracket - \rrbracket$ in the sense that any of the concrete instances of their conclusions follow from the concrete instances of their premises.
\end{theorem}

\begin{proof}
  (EdgeRename) and (BoxRename) are shown to follow from the other rules in Appendix~\ref{sec:app-renaming}. Proofs for the remaining rules are the subject of Appendix~\ref{sec:soundness}.
\end{proof}

A complete proof of soundness for (Induction) is left as future work (cf. Remark~\ref{rem:bang-logic} above).






\section{Examples}\label{sec:examples}

As mentioned in Section~\ref{sec:intro}, non-commutative nodes give us the ability to make recursive definitions of variable-arity generators in terms of fixed-arity generators of our theory. The induction principle in turn gives us the means to lift rules about fixed arity generators up to more powerful !-tensor rules. This section illustrates this concept, using a couple of simple examples that wouldn't be possible in the commutative setting.

\subsection{Monoids and tree-merging}

Suppose we take the theory of a monoid, i.e. the pair of generators \big(\input{Induc-Multiply.tikz},\begin{tikzpicture}[small]
	\begin{pgfonlayer}{nodelayer}
		\node [style=comm] (0) at (0, -0.25) {};
		\node [style=wire] (1) at (0, 0.5) {};
	\end{pgfonlayer}
	\begin{pgfonlayer}{edgelayer}
		\draw [style=directed] (0) to (1);
	\end{pgfonlayer}
\end{tikzpicture}\big) satisfying the associativity and unit laws. The diagrammatic theory is described by:
\begin{equation}\label{eqn:mon_theory}
  \Sigma_{Mon}:=\input{mon_sig.tikz},\qquad E_{Mon}:=\left\{\input{mon_eqns.tikz}\right\}
\end{equation}

We would like to define a new kind of node for $n$-fold trees of multiplications:
\ctikzfig{monoid_tree}
This can be done with a recursive definition:
\begin{equation}
	\input{Induc-SpiderBase.tikz} \qquad\qquad \input{Induc-SpiderRecurs.tikz}
\end{equation}
Note that this is well-defined, because unlike in the commutative case, an $n$-legged dot will unfold as a \textit{unique}, left-associated tree of multiplications, with a unit plugged into the leftmost position. We can verify that this acts as expected by expanding out the three input case using the definitions (the last step is a result of the unit law):
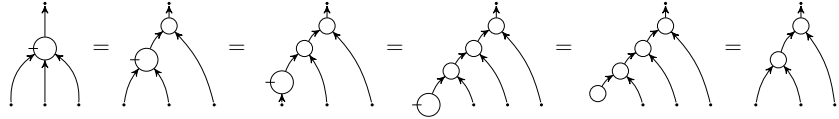
\ctikzfig{monoid_tree_expanding}

\begin{remark}
  Put another way, non-commutative !-boxes make such recursive definitions possible because they do not assume \textit{a priori} that the family of graphs generated by the definition are symmetric on their inputs/outputs. This need not be true, even in the case where all of the concrete generators are commutative. This limitation in the case of commutative !-boxes was highlighted in~\cite{MerryThesis}, where only a partial proof of the spider theorem for commutative Frobenius algebras could be done using (commutative) !-box induction.
\end{remark}

The first property we would like to prove about such trees is that any two connected trees merge to form bigger trees. As a !-box rule, it looks like this:
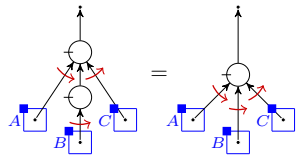
\ctikzfig{Induc-GenMergeTheorem}

To prove this we start by proving a lemma which does not involve the edges from !-box $C$:
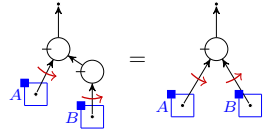
\begin{lemma}
  \quad\ctikzfig{Induc-MergeLemma}
\end{lemma}
\begin{proof}
  We can hit this lemma with induction on $B$ to break it into two cases:
  \begin{equation}\tag{base}
    \input{Induc-MergeBase-eq.tikz}
  \end{equation}
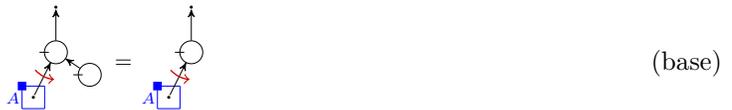
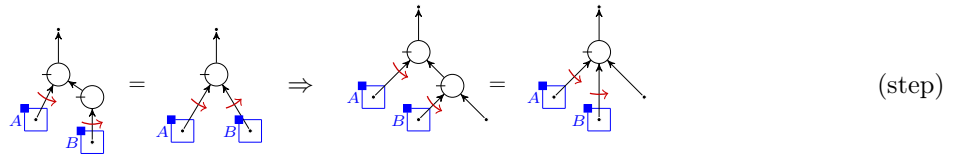
  \begin{equation}\tag{step}
    \input{Induc-MergeLemma.tikz} \ \ \Rightarrow\ \ \input{Induc-MergeStep-eq.tikz}
  \end{equation}
  ...each of which has a simple rewriting proof:
  \ctikzfig{Induc-MergeBase}
  \ctikzfig{Induc-MergeInduc}
  As pointed out above when we apply the induction hypothesis in step 4, the !-box $B$ must be `fixed' (i.e. we're not allowed to do any !-box operations on $B$ using $\Exp$, $\Kill$, $\Copy$, $\Drop$). This is because $B$ occurs free on both sides of the implication $G = H \Rightarrow \Exp_B(G = H)$. See~\cite{MerryThesis} for details.
\end{proof}

Now it is simple to check the general case.
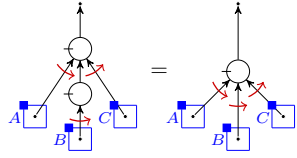
\begin{theorem}
  \quad\ctikzfig{Induc-GenMergeTheorem}
\end{theorem}
\begin{proof}
  This time we will apply induction on $C$. The base case is exactly the previous lemma, so we are only left to prove the step case, i.e.: 
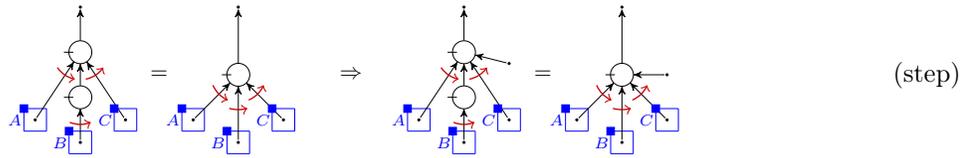
  \begin{equation}\tag{step}
    \input{Induc-MergeThmBase-eq.tikz}
  \end{equation}
  This is again a simple rewriting proof:
  \ctikzfig{Induc-MergeThmBase-step}
  Where we have used the inductive hypothesis in the second step.
\end{proof}

We can continue in this manner to prove rules like the merging rule (a.k.a. `spider rule') for commutative Frobenius algebras described in Section~\ref{sec:intro}. In fact, the proof goes through perfectly well in the non-commutative case, proving a merging rule for more general \textit{symmetric} Frobenius algebras, giving a purely diagrammatic characterisation of the normal forms described in~\cite{LaudaPfeiffer}.

As a corollary to our theorem we see that any arrangement of generators of the monoid can be combined in to one. In particular a mirror image tree is also equal to the variable arity node:
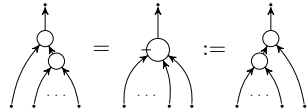
\ctikzfig{monoid_tree_mirror}

The rules of \eqref{eqn:mon_theory} can now be seen as concrete instances of our new !-box rule meaning we can reduce the diagrammatic theory of monoids from \eqref{eqn:mon_theory} to:

\[
\Sigma_{Mon} \ =\ \left\{ \ \input{mon_spider.tikz} \ \right\} \qquad
E_{Mon} \ =\ \left\{ \quad\input{Induc-GenMergeTheorem.tikz}, \quad
                        \input{ncspider_elim.tikz} \quad\right\}
\]

\subsection{Anti-homomorphisms}


Suppose that we introduce a new generator to the theory of monoids \eqref{eqn:mon_theory}, and impose two rules:
\[
\Sigma_{Antihom} \ =\ \Sigma_{Mon}\cup\ \left\{ \ \begin{tikzpicture}[small]
	\begin{pgfonlayer}{nodelayer}
		\node [style=anti] (0) at (0, 0) {};
		\node [style=wire] (1) at (0, 0.75) {};
		\node [style=wire] (2) at (0, -0.75) {};
	\end{pgfonlayer}
	\begin{pgfonlayer}{edgelayer}
		\draw [style=directed] (0) to (1);
		\draw [style=directed] (2) to (0);
	\end{pgfonlayer}
\end{tikzpicture} \ \right\} \qquad
E_{Antihom} \ =\ E_{Mon}\cup\ \left\{ \quad\input{Induc-Antihom-eqns.tikz} \quad\right\}
\]
This is called an \textit{anti-homomorphism}. We can imagine how this anti-homomorphism might interact with our recursively defined variable arity nodes:
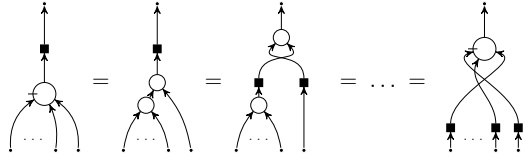
\ctikzfig{anti_hom_dots}

This can now be written down and proved rigorously using noncommutative !-boxes. Note that the arrows are in opposite directions on each sides of the following equation. This is what determines that the input edges reverse order.
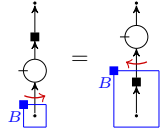
\begin{theorem}
  \[\input{anti_hom_box.tikz}\]
\end{theorem}
\begin{proof}
  Again we use !-box induction to reduce the problem to two easier tasks:
  \begin{equation}\tag{base}
    \input{anti_hom_box_base.tikz}
  \end{equation}
  \begin{equation}\tag{step}
    \input{anti_hom_box.tikz} \ \ \Rightarrow\ \ \input{anti_hom_box_step.tikz}
  \end{equation}
  The base case is true by definition from the diagrammatic equations of an anti-homomorphism (noting that variable arity nodes with no inputs are just units of the monoid). The Step case can then be proved using rewriting:
  \ctikzfig{anti_hom_box_step_proof}
  The third step is simply rearranging to make things easier to read. This is allowed as long as we keep track of which free edges are which (e.g by naming them).
\end{proof}

Since both defining equations of the anti-homomorphism are concrete instances of the variable arity case, the diagrammatic theory can now be written using only two generators and a much more powerful set of rules:
\[
\Sigma \ =\ \left\{ \ \input{mon_spider.tikz}, \quad 
                      \begin{tikzpicture}[small]
	\begin{pgfonlayer}{nodelayer}
		\node [style=anti] (0) at (0, 0) {};
		\node [style=wire] (1) at (0, 0.75) {};
		\node [style=wire] (2) at (0, -0.75) {};
	\end{pgfonlayer}
	\begin{pgfonlayer}{edgelayer}
		\draw [style=directed] (0) to (1);
		\draw [style=directed] (2) to (0);
	\end{pgfonlayer}
\end{tikzpicture} \ \right\} \qquad
E \ =\ \left\{ \quad\input{Induc-GenMergeTheorem.tikz}, \quad
                        \input{ncspider_elim.tikz}, \quad
                        \input{anti_hom_box.tikz}\quad\right\}
\]


\if\showproofs1

\appendix

\newpage

\bibliographystyle{eptcs}
\bibliography{bibfile}

\begin{thebibliography}{10}
\providecommand{\bibitemdeclare}[2]{}
\providecommand{\surnamestart}{}
\providecommand{\surnameend}{}
\providecommand{\urlprefix}{Available at }
\providecommand{\url}[1]{\texttt{#1}}
\providecommand{\href}[2]{\texttt{#2}}
\providecommand{\urlalt}[2]{\href{#1}{#2}}
\providecommand{\doi}[1]{doi:\urlalt{http://dx.doi.org/#1}{#1}}
\providecommand{\bibinfo}[2]{#2}

\bibitemdeclare{techreport}{Baez2014a}
\bibitem{Baez2014a}
\bibinfo{author}{John~C. \surnamestart Baez\surnameend} \&
  \bibinfo{author}{Jason \surnamestart Erbele\surnameend}
  (\bibinfo{year}{2014}): \emph{\bibinfo{title}{Categories in Control}}.
\newblock \bibinfo{type}{Technical Report},
  \bibinfo{institution}{arXiv:1405.6881}.

\bibitemdeclare{inproceedings}{SobocinskiSignal}
\bibitem{SobocinskiSignal}
\bibinfo{author}{F.~\surnamestart Bonchi\surnameend},
  \bibinfo{author}{P.~\surnamestart Sobocinski\surnameend} \&
  \bibinfo{author}{F.~\surnamestart Zanasi\surnameend} (\bibinfo{year}{2014}):
  \emph{\bibinfo{title}{A categorical semantics of signal flow graphs}}.
\newblock In: {\sl \bibinfo{booktitle}{CONCUR'14: Concurrency Theory.}}, {\sl
  \bibinfo{series}{Lecture Notes in Computer Science}} \bibinfo{volume}{8704},
  \bibinfo{publisher}{Springer}, pp. \bibinfo{pages}{435--450}.

\bibitemdeclare{article}{ContPhys}
\bibitem{ContPhys}
\bibinfo{author}{B.~\surnamestart Coecke\surnameend} (\bibinfo{year}{2009}):
  \emph{\bibinfo{title}{Quantum Picturalism}}.
\newblock {\sl \bibinfo{journal}{Contemporary Physics}} \bibinfo{volume}{51},
  pp. \bibinfo{pages}{59--83}.
\newblock \bibinfo{note}{{a}rXiv:0908.1787}.

\bibitemdeclare{inproceedings}{CD1}
\bibitem{CD1}
\bibinfo{author}{B.~\surnamestart Coecke\surnameend} \&
  \bibinfo{author}{R.~\surnamestart Duncan\surnameend} (\bibinfo{year}{2008}):
  \emph{\bibinfo{title}{Interacting quantum observables}}.
\newblock In: {\sl \bibinfo{booktitle}{Proceedings of the 37th International
  Colloquium on Automata, Languages and Programming (ICALP)}},
  \bibinfo{series}{Lecture Notes in Computer Science}.

\bibitemdeclare{inproceedings}{CDKZ}
\bibitem{CDKZ}
\bibinfo{author}{B.~\surnamestart Coecke\surnameend},
  \bibinfo{author}{R.~\surnamestart Duncan\surnameend},
  \bibinfo{author}{A.~\surnamestart Kissinger\surnameend} \&
  \bibinfo{author}{Q.~\surnamestart Wang\surnameend} (\bibinfo{year}{2012}):
  \emph{\bibinfo{title}{Strong complementarity and non-locality in categorical
  quantum mechanics}}.
\newblock In: {\sl \bibinfo{booktitle}{Proceedings of the 27th Annual IEEE
  Symposium on Logic in Computer Science (LICS)}}, \bibinfo{publisher}{IEEE
  Computer Society}.
\newblock \bibinfo{note}{ArXiv:1203.4988}.

\bibitemdeclare{article}{DixonDuncan2009}
\bibitem{DixonDuncan2009}
\bibinfo{author}{Lucas \surnamestart Dixon\surnameend} \& \bibinfo{author}{Ross
  \surnamestart Duncan\surnameend} (\bibinfo{year}{2009}):
  \emph{\bibinfo{title}{{Graphical Reasoning in Compact Closed Categories for
  Quantum Computation}}}.
\newblock {\sl \bibinfo{journal}{AMAI}}
  \bibinfo{volume}{56}(\bibinfo{number}{1}), p.~\bibinfo{pages}{20},
  \doi{10.1017/S0305004100074338}.

\bibitemdeclare{article}{DixonKissinger2010}
\bibitem{DixonKissinger2010}
\bibinfo{author}{Lucas \surnamestart Dixon\surnameend} \&
  \bibinfo{author}{Aleks \surnamestart Kissinger\surnameend}
  (\bibinfo{year}{2013}): \emph{\bibinfo{title}{Open-graphs and monoidal
  theories}}.
\newblock {\sl \bibinfo{journal}{Mathematical Structures in Computer Science}}
  \bibinfo{volume}{23}, pp. \bibinfo{pages}{308--359},
  \doi{10.1017/S0960129512000138}.
\newblock \bibinfo{note}{{a}rXiv:1007.3794v1~[cs.LO]}.

\bibitemdeclare{article}{JS}
\bibitem{JS}
\bibinfo{author}{Andre \surnamestart Joyal\surnameend} \& \bibinfo{author}{Ross
  \surnamestart Street\surnameend} (\bibinfo{year}{1991}):
  \emph{\bibinfo{title}{{The geometry of tensor calculus {I}}}}.
\newblock {\sl \bibinfo{journal}{Advances in Mathematics}}
  \bibinfo{volume}{88}, pp. \bibinfo{pages}{55--113},
  \doi{10.1016/0001-8708(91)90003-P}.

\bibitemdeclare{phdthesis}{DimitriDPhil}
\bibitem{DimitriDPhil}
\bibinfo{author}{D.~\surnamestart Kartsaklis\surnameend}
  (\bibinfo{year}{2014}): \emph{\bibinfo{title}{Compositional Distributional
  Semantics with Compact Closed Categories and Frobenius Algebras}}.
\newblock Ph.D. thesis, \bibinfo{school}{University of Oxford}.

\bibitemdeclare{phdthesis}{KissingerThesis}
\bibitem{KissingerThesis}
\bibinfo{author}{Aleks \surnamestart Kissinger\surnameend}
  (\bibinfo{year}{2011}): \emph{\bibinfo{title}{{Pictures of Processes:
  Automated Graph Rewriting for Monoidal Categories and Applications to Quantum
  Computing}}}.
\newblock Ph.D. thesis, \bibinfo{school}{University of Oxford}.
\newblock \bibinfo{note}{{a}rXiv:1203.0202 [math.CT]}.

\bibitemdeclare{incollection}{KissingerATS}
\bibitem{KissingerATS}
\bibinfo{author}{Aleks \surnamestart Kissinger\surnameend}
  (\bibinfo{year}{2014}): \emph{\bibinfo{title}{Abstract Tensor Systems as
  Monoidal Categories}}.
\newblock In \bibinfo{editor}{C~\surnamestart Casadio\surnameend},
  \bibinfo{editor}{B~\surnamestart Coecke\surnameend},
  \bibinfo{editor}{M~\surnamestart Moortgat\surnameend} \&
  \bibinfo{editor}{P~\surnamestart Scott\surnameend}, editors: {\sl
  \bibinfo{booktitle}{Categories and Types in Logic, Language, and Physics:
  Festschrift on the occasion of Jim Lambek's 90th birthday}}, {\sl
  \bibinfo{series}{Lecture Notes in Computer Science}} \bibinfo{volume}{8222},
  \bibinfo{publisher}{Springer}, \doi{10.1007/978-3-642-54789-8\_13}.
\newblock \bibinfo{note}{{arXiv:1308.3586 [math.CT]}}.

\bibitemdeclare{inproceedings}{PatternGraph}
\bibitem{PatternGraph}
\bibinfo{author}{Aleks \surnamestart Kissinger\surnameend},
  \bibinfo{author}{Alex \surnamestart Merry\surnameend} \&
  \bibinfo{author}{Matvey \surnamestart Soloviev\surnameend}
  (\bibinfo{year}{2012}): \emph{\bibinfo{title}{Pattern Graph Rewrite
  Systems}}.
\newblock In: {\sl \bibinfo{booktitle}{Proceedings of DCM 2012}}, {\sl
  \bibinfo{series}{EPTCS}} \bibinfo{volume}{143}, \doi{10.4204/EPTCS.143.5}.
\newblock \bibinfo{note}{{a}rXiv:1204.6695 [math.CT]}.

\bibitemdeclare{misc}{Quantomatic}
\bibitem{Quantomatic}
\bibinfo{author}{Aleks \surnamestart Kissinger\surnameend},
  \bibinfo{author}{Alexander \surnamestart Merry\surnameend},
  \bibinfo{author}{Lucas \surnamestart Dixon\surnameend}, \bibinfo{author}{Ross
  \surnamestart Duncan\surnameend}, \bibinfo{author}{Matvey \surnamestart
  Soloviev\surnameend} \& \bibinfo{author}{Benjamin \surnamestart
  Frot\surnameend} (\bibinfo{year}{2011}):
  \emph{\bibinfo{title}{{Quantomatic}}}.
\newblock \bibinfo{howpublished}{https://sites.google.com/site/quantomatic/}.

\bibitemdeclare{inproceedings}{NoncommBB}
\bibitem{NoncommBB}
\bibinfo{author}{Aleks \surnamestart Kissinger\surnameend} \&
  \bibinfo{author}{David \surnamestart Quick\surnameend}
  (\bibinfo{year}{2014}): \emph{\bibinfo{title}{Tensors, !-graphs, and
  non-commutative quantum structures}}.
\newblock In: {\sl \bibinfo{booktitle}{Proceedings of the 11th workshop on
  Quantum Physics and Logic, {QPL} 2014, Kyoto, Japan, 4-6th June 2014.}}, pp.
  \bibinfo{pages}{56--67}, \doi{10.4204/EPTCS.172.5}.
\newblock \bibinfo{note}{{a}rXiv:1412.8552 [cs.LO]}.

\bibitemdeclare{article}{LackAdh2005}
\bibitem{LackAdh2005}
\bibinfo{author}{Stephen \surnamestart Lack\surnameend} \&
  \bibinfo{author}{Pawel \surnamestart Sobocinski\surnameend}
  (\bibinfo{year}{2005}): \emph{\bibinfo{title}{{Adhesive and quasiadhesive
  categories}}}.
\newblock {\sl \bibinfo{journal}{Theoretical Informatics and Applications}}
  \bibinfo{volume}{39}(\bibinfo{number}{2}), pp. \bibinfo{pages}{522--546},
  \doi{10.1051/ita:2005028}.

\bibitemdeclare{article}{LaudaPfeiffer}
\bibitem{LaudaPfeiffer}
\bibinfo{author}{Aaron~D. \surnamestart Lauda\surnameend} \&
  \bibinfo{author}{Hendryk \surnamestart Pfeiffer\surnameend}
  (\bibinfo{year}{2008}): \emph{\bibinfo{title}{Open-closed strings:
  Two-dimensional extended TQFTs and Frobenius algebras}}.
\newblock {\sl \bibinfo{journal}{Topology Appl.}}
  \bibinfo{volume}{155}(\bibinfo{number}{7}), pp. \bibinfo{pages}{623--666},
  \doi{10.1016/j.topol.2007.11.005}.

\bibitemdeclare{phdthesis}{MerryThesis}
\bibitem{MerryThesis}
\bibinfo{author}{Alexander \surnamestart Merry\surnameend}
  (\bibinfo{year}{2014}): \emph{\bibinfo{title}{Reasoning with !-Graphs}}.
\newblock Ph.D. thesis, \bibinfo{school}{University of Oxford}.

\bibitemdeclare{incollection}{Penrose1971}
\bibitem{Penrose1971}
\bibinfo{author}{R.~\surnamestart Penrose\surnameend} (\bibinfo{year}{1971}):
  \emph{\bibinfo{title}{Applications of negative dimensional tensors}}.
\newblock In: {\sl \bibinfo{booktitle}{Combinatorial Mathematics and its
  Applications}}, \bibinfo{publisher}{Academic Press}, pp.
  \bibinfo{pages}{221--244}.

\end{thebibliography}

\newpage

\section{Additional Proofs}

\subsection{Preservation of !-tensor expressions by !-box operations}\label{app:ops_create_tensors}

  We wish to prove Theorem~\ref{thm:ops_create_tensors} which states that the application of a !-box operation to a !-tensor expression results in a !-tensor expression. First we present a Lemma describing how !-box operations affect contexts.

  \begin{lemma}
    \label{lem:contexts}
    If $\ectx_G(a)=[E_1,\ldots,E_n]$, $\nctx_G(a)=[N_1,\ldots,N_m]$ then the following shows contexts affected by operations (writing $B'$ for $\fr(B)$):
    \begin{table}[ht]
    \renewcommand{\arraystretch}{1.2}
    \begin{tabular} {|r|l|c|c|} \hline
                    &            & $\ectx$ & $\nctx$ \\ \hline
      $\Drop_{E_i}$ &        $a$ & $[E_1,\ldots,E_{i-1},E_{i+1},\ldots,E_n]$ &  \\ \hline
      $\Drop_{N_i}$ &        $a$ &  & $[N_1,\ldots,N_{i-1},N_{i+1},\ldots,N_m]$ \\ \hline
      $\Exp_{E_i}$  &   $\!\fr(a)\!$ & $[E_1',\ldots,E_{i-1}',E_{i+1},\ldots,E_n]$ &  \\ \hline
      $\Exp_{N_i}$  &   $\!\fr(a)\!$ & $[E_1',\ldots,E_n']$ & $[N_1',\ldots,N_{i-1}',N_{i+1},\ldots,N_m]$ \\ \hline
      $\Copy_{E_i}$ &   $\!\fr(a)\!$ & $[E_1',\ldots,E_i',E_{i+1},\ldots,E_n]$ & \\ \hline
      $\Copy_{N_i}$ &   $\!\fr(a)\!$ & $[E_1',\ldots,E_n']$ & $[N_1',\ldots,N_i',N_{i+1},\ldots,N_m]$ \\ \hline
    \end{tabular}
    \end{table}\\
    Note that any edges remaining after a $\Kill$ operation have not had their contexts affected.
  \end{lemma}

  We omit the proof, which follows mechanically from the definitions of each respective operation.
  By definition, the results of !-box operations are !-pretensors. To additionally show that $\Op_A(G)$ is a valid !-tensor expression, i.e. that $\Op_A(G) \in \graphterms$, we need to show that the !-tensor conditions (Definition~\ref{def:tensor_conditions}) still hold. In each case the conditions F1-2 are trivial since new edges/!-boxes have new names created by a freshness function. Hence we will only check conditions C1-3 for each of our !-box operations. 

  \begin{theorem}\label{thm:kill_well_defined}
    $G\in\graphterms \implies \Kill_A(G)\in\graphterms$.
  \end{theorem}
  \begin{proof}
    Using Lemma~\ref{lem:contexts}, we check C1-3:
    \begin{itemize}
        \item[C1:] If $a\in\Edges(\Kill_A(G))$ then contexts were not affected by $\Kill_A$ so the same condition holds.
        \item[C2:] Suppose we have $a\in\Edges(\Kill_A(G))$ with edge context $[E_1,\ldots,E_n]$.
        We then have $a\in \Edges(G)$ with edge context $[E_1,\ldots,E_n]$, hence $E_1\prec_G\ldots\prec_G E_n$ (by C2 on $G$), with no $E_i$ nested inside $A$. Then,
        $E_1\prec_{\Kill_A(G)}\ldots\prec_{\Kill_A(G)} E_n$.
        \item[C3:] If $\eout{a},\ein{a}\in\Edges(\Kill_A(G))$ then $\eout{a},\ein{a}\in\Edges(G)$ and $\ectx, \nctx$ were not affected by $\Kill_A$ so the condition still holds.
    \end{itemize}
  \end{proof}

  \begin{theorem}\label{thm:drop_well_defined}
    $G\in\graphterms \implies \Drop_A(G)\in\graphterms$.
  \end{theorem}
  \begin{proof}
    Using Lemma~\ref{lem:contexts}, we check C1-3:
    \begin{itemize}
        \item[C1:] Again trivial since $\ectx$ and $\nctx$ have only lost !-boxes.
        \item[C2:] If $a\in\Edges(\Drop_A(G))$ has edge context $[E_1,\ldots,E_n]$
          then $a\in \Edges(G)$ could have the same edge context or contain $A$, i.e $[E_1,\ldots,A,\ldots,E_n]$, in which case nesting in $G$ would be $E_1\prec_G\ldots\prec_G A \prec_G\ldots\prec_G E_n$.
          In either case, the nesting in $\Drop_A(G)$ is $E_1\prec_{\Drop_A(G)}\ldots\prec_{\Drop_A(G)} E_n$ since $A$ is removed.
        \item[C3:] If $\eout{a},\ein{a}\in\Edges(\Drop_A(G))$ then $\eout{a},\ein{a}\in\Edges(G)$ and $\ectx, \nctx$ only lost the !-box $A$ so the condition still holds by removing $A$ from $es, bs$.
    \end{itemize}
  \end{proof}

  \begin{theorem}\label{thm:copy_well_defined}
    $G\in\graphterms \implies \Copy_A(G)\in\graphterms$.
  \end{theorem}
  \begin{proof}
    Using Lemma~\ref{lem:contexts}, we check C1-3:
    \begin{itemize}
      \item[C1:] Edges in $\Copy_A(G)$ are either edges from $G$ or fresh names for edges in $G$. For the former $\ectx, \nctx$ have not been changed so the condition holds. For the latter we know from Lemma~\ref{lem:contexts} that !-boxes in $\ectx, \nctx$ have only been replaced by fresh versions of themselves hence are still distinct.
      \item[C2:] For $a\in\Edges(G)$ (writing $B'$ for $\fr(B)$):
        \begin{itemize}
          \item[$\bullet$] For $a$ in $G$ we have $\ectx_{\Copy_A(G)}(a)=\ectx_G(a)$ and $E_i\prec_G E_{i+1} \implies E_i\prec_{\Copy_A(G)} E_{i+1}$ so the condition holds.
          \item[$\bullet$] $\ectx_{\Copy_A(G)}(\fr(a)) = [E_1',\ldots,E_i',E_{i+1},\ldots,E_n]$ where $\ectx_G(a)=[E_1,\ldots,E_n]$ and $A=E_i$.
          From the definition of $\Copy_A$ we see $E_i'$ ($=A'$) must have been created inside $E_{i+1}$. For the other !-boxes, it is clear from $\ectx_G(a)$ that
          $E_j'\prec_{\Copy_A(G)} E_{j+1}'$ for $j<i$ and
          $E_j\prec_{\Copy_A(G)} E_{j+1}$ for $j>i$.
        \end{itemize}
      \item[C3:] For $\eout{a},\ein{a}\in\Edges(\Copy_A(G))$ where $\eout{a},\ein{a}\in\Edges(G)$, $\ectx$ and $\nctx$ are not affected by $\Copy_A$ so the condition still holds.

      For $\fr(\eout{a}),\fr(\ein{a})\in\Edges(\Copy_A(G))$ where $\eout{a},\ein{a}\in\Edges(G)$, there exist !-boxes $E_i,N_i \in \boxnames$ such that: \\
      $[E_1,\ldots,E_n].\nctx_G(\ein{a})=\ectx_G(\eout{a}).[N_1,\ldots,N_m]$ and
      $[E_1,\ldots,E_n].\nctx_G(\eout{a})=\ectx_G(\ein{a}).[N_1,\ldots,N_m]$.
      
      Since the edge was copied, $A$ must be in the contexts of $\eout{a}$ and $\ein{a}$. We need to show that there exist $es, bs$ such that:
      \[ es.\nctx_G(\ein{a}) = \ectx_G(\eout{a}).bs
       \quad \textrm{and} \quad
        es.\nctx_G(\eout{a}) = \ectx_G(\ein{a}).bs \]
      We do this by considering 4 cases:
      \begin{itemize}
        \item[$\bullet$] If $A\in \ectx_G(\ein{a})\cap \ectx_G(\eout{a})$ then $A=E_i$ for some $i$, and the condition holds by letting $es:=[E_1',\ldots,E_i',E_{i+1},\ldots,E_n]$ and $bs:=[N_1,\ldots,N_m]$.
        \item[$\bullet$] Similarly, if $A\in \nctx_G(\ein{a})\cap \nctx_G(\eout{a})$ then $A=N_i$ for some $i$, and the condition holds by letting $es:=[E_1',\ldots,E_m']$ and $bs:=[N_1',\ldots,N_i',N_{i+1},\ldots,N_n]$.
        \item[$\bullet$] If $A\in \ectx_G(\ein{a})\cap \nctx_G(\eout{a})$, the condition holds with $es:=[E_1',\ldots,E_m']$ and $bs:=[N_1,\ldots,N_m]$.
        \item[$\bullet$] Otherwise $A\in \nctx_G(\ein{a})\cap \ectx_G(\eout{a})$, in which case the condition holds with $es:=[E_1',\ldots,E_m']$ and $bs:=[N_1,\ldots,N_m]$.
      \end{itemize}
    \end{itemize}
  \end{proof}

  \begin{theorem}\label{thm:exp_well_defined}
    $G\in\graphterms \implies \Exp_A(G)\in\graphterms$.
  \end{theorem}
  \begin{proof}
    We note that $\Exp_{A,\fr}=\Drop_{\fr(A)}\circ\Copy_{A,\fr}$ so this case follows from the previous cases.
  \end{proof}

  \subsection{Preservation of !-tensor equivalence by !-box operations}\label{app:ops_pres_equiv}

  \begin{theorem}
    Let $\fr$ be a freshness function for the !-tensor expressions $G, H$. Then $G\equiv H$ implies $\Op_{B,\fr}(G) \equiv \Op_{B,\fr}(H)$.
  \end{theorem}
  \begin{proof}
    We need to check our enforced equivalences still hold after $\Op_B$. It is clear from the definitions of $\Op_B(G H), \Op_B(e f), \Op_B(1), \Op_B(\epsilon)$ that associativity/commutativity/unit conditions are preserved. We check the other cases:
    \begin{itemize}
      \item $\term{\Op_B([\epsilon>A)}\equiv\term{[\Op_B(\epsilon)>A}\equiv\term{[\epsilon>A}\equiv\epsilon\equiv\ldots\equiv\term{\Op_B(<\epsilon]A)}$ \quad if $A\not= B$ \\
      When $A=B$ we need to check each operation individually:
      \subitem $\term{\Kill_B([\epsilon>B)}\equiv\term{\epsilon}\equiv\term{\Kill_B(<\epsilon]B)}$
      \subitem $\term{\Exp_{B,\fr}([\epsilon>B)}\equiv\term{[\epsilon>B\epsilon}\equiv\term{\epsilon}\equiv\term{\epsilon<\epsilon]B}\equiv\term{\Exp_{B,\fr}(<\epsilon]B)}$
      \subitem $\term{\Drop_B([\epsilon>B)}\equiv\term{\epsilon}\equiv\term{\Drop_B(<\epsilon]B)}$
      \subitem $\term{\Copy_{B,\fr}([\epsilon>B)}\equiv\term{[\epsilon>B[\epsilon>{\fr(B)}}\equiv\term{\epsilon}\equiv\term{<\epsilon]{\fr(B)}<\epsilon]B}\equiv\term{\Copy_{B,\fr}(<\epsilon]B)}$
      \item We need to check $\term{G[K_n...[K_1 1.+b-a.]{B_1}...]{B_n} \equiv G[-b \mapsto -a] [K_n...[K_1]{B_1}...]{B_n}}$ is preserved given free name $\term{-b}$ in $G$.  \\
      Looking at the left hand side we see $\ectx(\term{+b})=[]$ and $\nctx(\term{+b})=[B_1,\ldots,B_n]$. \\
      To be well-defined (C2) then ensures that $\nctx(\term{-b})=[]$ and $\ectx(\term{-b})=[B_1,\ldots,B_n]$. \\
      If $B\not\in[B_1,\ldots,B_n]$ then neither $\term{+b}$ nor $\term{-b}$ is affected by $\Op_B$ so we get:
      \begin{align*}
        \term{\Op_B(G[K_n...[K_1 1.+b-a.]{B_1}...]{B_n})} &\equiv \term{\Op_B(G)[\Op_B(K_n)...[\Op_B(K_1) 1.+b-a.]{B_1}...]{B_n})} \\
            &\equiv \term{\Op_B(G)[-b \mapsto -a] [\Op_B(K_n)...[\Op_B(K_1)]{B_1}...]{B_n})} \\
            &\equiv \term{\Op_B(G[-b \mapsto -a] )\Op_B([K_n...[K_1]{B_1}...]{B_n}))} \\
            &\equiv \term{\Op_B(G[-b \mapsto -a] [K_n...[K_1]{B_1}...]{B_n})}
      \end{align*}
      Otherwise $B=B_i$ for some $i\leq n$. Then each operation needs to be checked individually:
      \begin{itemize}
        \item
        $\term{\Kill_B(G[K_n...[K_1 1.+b-a.]{B_1}...]{B_n})}$
        \vspace{-9pt}
        \begin{flalign*}
          \quad &\equiv \term{\Kill_B(G)[\Kill_B(K_n)...[K_{i+1} 1]{B_{i+1}}...]{B_n}} & \\
                &\equiv \term{\Kill_B(G)\Kill_B([K_n...[K_1]{B_1}...]{B_n})} & \\
                &\equiv \term{\Kill_B(G[-b \mapsto -a] )\Kill_B([K_n...[K_1]{B_1}...]{B_n})} & \\
                &\equiv \term{\Kill_B(G[-b \mapsto -a] [K_n...[K_1]{B_1}...]{B_n})} &
        \end{flalign*}
        Where the second to last equivalence is true since $\Kill_B$ will delete the edge in to $G$ whether it is named $\term{-b}$ or $\term{-a}$.
        \item
        $\term{\Drop_B(G[K_n...[K_1 1.+b-a.]{B_1}...]{B_n})}$
        \vspace{-9pt}
        \begin{flalign*}
          \quad &\equiv \term{\Drop_B(G)[\Drop_B(K_n)...[\Drop_B(K_{i+1})K_i[...[K_1 1.+b-a.]{B_1}...]{B_{i-1}}]{B_{i+1}}...]{B_n}} & \\
                &\equiv \term{\Drop_B(G)[-b \mapsto -a] [\Drop_B(K_n)...[\Drop_B(K_{i+1})K_i[...[K_1]{B_1}...]{B_{i-1}}]{B_{i+1}}...]{B_n}} & \\
                &\equiv \term{\Drop_B(G[-b \mapsto -a] )\Drop_B([K_n...[K_1]{B_1}...]{B_n})} & \\
                &\equiv \term{\Drop_B(G[-b \mapsto -a] [K_n...[K_1]{B_1}...]{B_n})}
        \end{flalign*}
        Where the second to last equivalence is true since $\Drop_B$ will act the same whether our edge is named $\term{-b}$ or $\term{-a}$.
        \item Abbreviating $\Copy_{B,\fr}$ as $C$ and $\fr(x)$ as $x'$:\\
        $\term{C(G[K_n...[K_1 1.+b-a.]{B_1}...]{B_n})}$
        \vspace{-9pt}
        \begin{flalign*}
          \quad &\equiv \term{C(G)[C(K_n)...[K_i...[K_1 1.+b-a.]{B_1}]{B_i}[K_i'[...[K_1' 1.+b'-a'.]{B_1'}]{B_i'}...]{B_n}} & \\
                &\equiv \term{C(G)[-b \mapsto -a] [-b' \mapsto -a'] [C(K_n)...[K_i...[K_1]{B_1}...]{B_i}[K_i'...[K_1']{B_1'}...]{B_i'}...]{B_n}} & \\
                &\equiv \term{C(G[-b \mapsto -a] )C([K_n...[K_1]{B_1}...]{B_n})} & \\
                &\equiv \term{C(G[-b \mapsto -a] [K_n...[K_1]{B_1}...]{B_n})}
        \end{flalign*}
        Where the second to last equivalence is true since moving a rename to after $\Copy_{B,\fr}$ you have to do both the original rename and the fresh version.
        \item $\Exp_B$ follows from the fact that $\Exp_{B,\fr}=\Drop_{\fr(B)}\circ\Copy_{B,\fr}$.
      \end{itemize}
      \item The proof for $\term{H[K_n...[K_1 1.+a-b.]{B_1}...]{B_n} \equiv H[+b \mapsto +a]}$ is similar.
    \end{itemize}
  \end{proof}

\subsection{Renaming from existing rules}\label{sec:app-renaming}

As mentioned in Section~\ref{sec:box-reasoning}, (EdgeRename) follows from (Prod) and (Weaken).

\begin{lemma}
  If $\ctx_G(\eout{a})=[B,\ldots]$ then $G[\eout{a}\mapsto\eout{b}]\equiv\Wk{B}{\term{1.+b-a.}}(G)$.
\end{lemma}
\begin{proof}
  This follows immediately from !-tensor equivalence \eqref{eqn:contract_equiv}, so we just need to check that $G':=\Wk{B}{\term{1.+b-a.}}(G)$ is always well-defined. All of the !-tensor conditions except C3 are trivially satisfied, so it suffices to show C3.
  Note $\ectx_{G'}(\term{-a})=[]$ and $\nctx_{G'}(\term{-a})=[B,\ldots]=\ctx_G(\eout{a})$.
  Letting $es:=[]$ and $ns:=\nctx_G(\eout{a})$ we have 
  $es.\nctx_{G'}(\term{+a}) = \nctx_G(\eout{a}) = \ectx_{G'}(\term{-a}).ns$ and
  $es.\nctx_{G'}(\term{-a}) = \ctx_G(\eout{a}) = \ectx_{G'}(\term{+a}).ns$, which completes the proof.
\end{proof}

Similarly we can check that $G[\ein{a}\mapsto\ein{b}]\equiv\Wk{B}{\term{1.+a-b.}}(G)$ where $\ctx_G(\ein{a})=[B,\ldots]$.

\begin{theorem}
  \quad
  \[
    \AxiomC{$G = H$}
    \RightLabel{\rm\scriptsize(EdgeRename)}
    \UnaryInfC{$G[a\mapsto b] = H[a\mapsto b]$}
    \DisplayProof
  \]
  Where the edge $a$ is free in $G$ (and therefore also in $H$). 
\end{theorem}
\begin{proof}
  Without loss of generality we take the case where $\term{+a}$ is free in $G,H$. \\
  If $\ctx_G(\term{+a})=[]$ then the proof can be derived from (Prod):
  \[
    \AxiomC{$G = H$}
    \RightLabel{\scriptsize(Prod)}
    \UnaryInfC{$\term{G1.+b-a.} = \term{H1.+b-a.}$}
    \RightLabel{\scriptsize($\equiv$)}
    \UnaryInfC{$G[a\mapsto b] = H[a\mapsto b]$}
    \DisplayProof
  \]
  Otherwise let $\ctx_G(\term{+a})=[B,\ldots]$ then by the previous Lemma using (Weaken):
  \[
    \AxiomC{$G = H$}
    \RightLabel{\scriptsize(Weaken)}
    \UnaryInfC{$\Wk{B}{\term{1.+b-a.}}(G) = \Wk{B}{\term{1.+b-a.}}(H)$}
    \RightLabel{\scriptsize($\equiv$)}
    \UnaryInfC{$G[a\mapsto b] = H[a\mapsto b]$}
    \DisplayProof
  \]
\end{proof}

Using ($\Copy$) and ($\Kill$) we can also prove (BoxRename).


\begin{theorem}
  \quad
  \[
    \AxiomC{$G = H$}
    \RightLabel{\rm\scriptsize(BoxRename)}
    \UnaryInfC{$G[A\mapsto B] = H[A\mapsto B]$}
    \DisplayProof
  \]
  where $A$ is a !-box in $G=H$. 
\end{theorem}
\begin{proof}
  Given $G$ if we apply $\Kill_A\circ\Copy_{A,\fr}$ we have $G$ with everything inside (and including) $A$ renamed by $\fr$. i.e. $\Kill_A\circ\Copy_{A,\fr}=G[\fr|_{A}]$ where $\fr|_{A}$ is $\fr$ restricted to contents of !-box $A$ in $G$. We will use this operation twice to rename everything the way we want. \\
  Let $\fr$ be a freshness function for $G$ s.t. $\fr(A)=A'$ for some fresh name $A'$ and $\fr(A')=B, \fr(B)=A$. \\
  Let $\fr'(C):=\begin{cases}
    B & C=A' \\
    A & C=B \\
    A' & C=A \\
    \fr^{-1} & \text{otherwise} \\
  \end{cases}$ \\
  So applying these two operations on the contents of $A$ results in $A\mapsto A'\mapsto B$, and for all other names in $A$ we have identity: $C\mapsto \fr(C)\mapsto C$. \\
  So we can prove:
  \[
    \AxiomC{$G = H$}
    \RightLabel{\scriptsize($\Copy_{A,\fr}$)}
    \UnaryInfC{$\Copy_{A,\fr}(G) = \Copy_{A,\fr}(H)$}
    \RightLabel{\scriptsize($\Kill_A$)}
    \UnaryInfC{$\Kill_A\circ\Copy_{A,\fr}(G) = \Kill_A\circ\Copy_{A,\fr}(H)$}
    \RightLabel{\scriptsize($\equiv$)}
    \UnaryInfC{$G[\fr|_{A}] = H[\fr|_{A}]$}
    \RightLabel{\scriptsize($\Copy_{A',\fr'}$)}
    \UnaryInfC{$\Copy_{A',\fr'}(G[\fr|_{A}]) = \Copy_{A',\fr'}(H[\fr|_{A}])$}
    \RightLabel{\scriptsize($\Kill_{A'}$)}
    \UnaryInfC{$\Kill_{A'}\circ\Copy_{A',\fr'}(G[\fr|_{A}]) = \Kill_{A'}\circ\Copy_{A',\fr'}(H[\fr|_{A}])$}
    \RightLabel{\scriptsize($\equiv$)}
    \UnaryInfC{$G[\fr|_{A}][\fr'|_{A'}] = H[\fr|_{A}][\fr'|_{A'}]$}
    \RightLabel{\scriptsize($\equiv$)}
    \UnaryInfC{$G[A\mapsto B] = H[A\mapsto B]$}
    \DisplayProof
  \]
\end{proof}

\subsection{Soundness of !-box inference rules}\label{sec:soundness}

In Section~\ref{sec:box-reasoning}, we gave an interpretation $\llbracket-\rrbracket$ of !-tensor equations in terms of sets of concrete equations, which can in turn be modelled in a category $\mathcal C$. We now show that our !-box rules are sound with respect to this interpretation. In each case we need to check that any concrete instance of the conclusion is implied by the concrete instances of the premises.

\begin{theorem}
  $(\Exp_B)$ and $(\Kill_B)$ are sound with respect to $\llbracket-\rrbracket$.
\end{theorem}
\begin{proof}
  Given $i\in\inst{\Exp_B(G=H)}$, by definition we have $i(\Exp_B(G=H))$ is a concrete tensor and so $i\circ\Exp_B\in\inst{G=H}$, so $\llbracket \Exp_B(G=H) \rrbracket \subseteq \llbracket G=H \rrbracket$. The $\Kill_B$ case is similar.
\end{proof}

\begin{theorem}
  $(\Copy_B)$ is sound with respect to $\llbracket-\rrbracket$.
\end{theorem}
\begin{proof}
  Suppose $j$ is a combination of $\Kill$, $\Exp$ and $\Copy$ operations such that $j(G=H)$ is a concrete tensor equation. We wish to show that $j$ can be rewritten to an instantiation, that is, a composition of only $\Kill$ and $\Exp$ operations. We do this by repeatedly applying the following rewriting procedure to $j$: \\
  If $j$ has no $\Copy$ operations then we are done, else suppose $\Copy_B$ is the left-most $\Copy$ operation, so $j=i\circ\Copy_B\circ j'$ where $i$ is an instantiation of $\Copy_B\circ j'(G=H)$. Now we have two cases:
  \begin{itemize}
    \item $B$ has a parent !-box in $\Copy_B\circ j'(G=H)$. Let $A$ be the top level !-box containing $B$. By Theorem~\ref{thm:top_op_first} we can write $i=i'\circ\KE^n_A$ so that:
    \begin{align*}
      j &= i\circ\Copy_B\circ j' \\
        &= i'\circ\KE^N_A\circ\Copy_B\circ j' \\
        &= i'\circ\Copy_{B_1}\circ\ldots\circ\Copy_{B_n}\circ\KE^n_A\circ j'
    \end{align*}
    Where $B_1,\ldots B_n$ are the copies of $B$ created by $\KE^n_A$. \\
    The final step here is easily checked from the recursive definitions of !-box operations. \\
    Now we can restart the procedure on $i'\circ\Copy_{B_1}\circ\ldots\circ\Copy_{B_n}\circ\KE^n_A\circ j'$.
    \item $B$ has no parent !-box in $\Copy_B\circ j'(G=H)$. Then $\fr(B)$ also has no parent !-box in $\Copy_B\circ j'(G=H)$ where $\fr$ is the freshness function from $\Copy_B$. By Theorem~\ref{thm:top_op_first} we can write $i=i'\circ\KE^n_{\fr(B)}$ so that:
    \begin{align*}
      j &= i\circ\Copy_{B,\fr}\circ j' \\
        &= i'\circ\KE^N_{\fr(B)}\circ\Copy_{B,\fr}\circ j' \\
        &= i'\circ\Exp^n_B\circ j'
    \end{align*}
    Where the freshness functions for $\Exp^n_B$ are chosen to give the same names as those from $\KE^N_{\fr(B)}$. \\
    Now we can restart the procedure on $i'\circ\Exp^n_{\fr(B)}\circ j'$
  \end{itemize}
  Given an instantiation $i\in\inst{\Copy_B(G=H)}$ we use this procedure to rewrite $i\circ\Copy_B$ until it is an instantiation of $G=H$ so that it follows from $\sem{G=H}$. \\
  To see that this process terminates, note that each iteration either eliminates a copy operation at depth $0$, or replaces a copy operation at depth $k$ with operations at depth $k-1$.
  \end{proof}

For soundness of (Weaken) we first present a Lemma about how !-box operations affect weakened !-tensors:

\begin{lemma}\label{lem:weaken_reorder}
  If $A$ is nested inside $B$ then the following demonstrate the affect of applying !-box operations to $\Wk{A}{K}(G)$. Operations on !-boxes not nested in or containing $A$ commute with $\Wk{A}{K}$.
  \begin{itemize}
    \item $\Kill_B(\Wk{A}{K}(G))=\Kill_B(G)$
    \item $\Exp_{B,\fr}(\Wk{A}{K}(G))=\Wk{A}{K}\circ\Wk{\fr(A)}{\fr(K)}(\Exp_{B,\fr}(G))$
    \item $\Kill_A(\Wk{A}{K}(G))=\Kill_A(G)$
    \item $\Exp_{A,\fr}(\Wk{A}{K}(G))=\begin{cases}
      \Wk{A}{K}\circ\Wk{C}{\fr(K)}(\Exp_{A,\fr}(G)) & A\prec_G C \\
      \Wk{A}{K}(\Exp_{A,\fr}(G)\fr(K)) & A \text{ top level}
    \end{cases}$
  \end{itemize}
\end{lemma}
\begin{proof}
  Proof is by structural induction on $G$.
\end{proof}

\begin{theorem}
  (Weaken) is sound with respect to $\llbracket-\rrbracket$.
\end{theorem}
\begin{proof}
  We prove that the set of concrete instances $\sem{G=H}$ imply all of the instances $\sem{\Wk{A}{K}G=\Wk{A}{K}H}$ by induction on the depth at which $A$ is nested in $G=H$.
  \begin{itemize}
    \item For the base case $A$ has no nesting so is a top level !-box. Assume $\sem{G = H}$. Then, for any $n$, $\sem{\KE^n_A(G) = \KE^n_A(H)}$ all hold. Let $K_1$ to $K_n$ be !-tensors, obtained from $K$ by applications of freshness functions from $\KE^n_A$ to $K$. Invoking the product rule yields:
    \[ \sem{K_n\ldots K_1\KE^n_A(G) = K_n\ldots K_1\KE^n_A(H)} \]
    which then implies
    $\sem{\KE^n_A (\Wk{A}{K}G) = \KE^n_A (\Wk{A}{K}H)}$
    by Lemma~\ref{lem:weaken_reorder}. Since this holds for all $n$, $\sem{\Wk{A}{K}G = \Wk{A}{K}H}$ follows from Theorem~\ref{thm:top_op_first}.
    \item For the step case, suppose the theorem holds for all !-boxes nested less deeply than a fixed !-box $A$, where $A\prec_G \ldots \prec_G B$ with $B$ top level. Again, we start by expanding the top-level !-box, for some $n$:
    \[ \sem{G = H}\Rightarrow\sem{\KE^n_B(G) = \KE^n_B(H)} \]
    Let $A_i$ and $K_i$ be $n$ copies of $A$ and $K$ respectively, using the freshness functions from $\KE^n_B$. Each $A_i$ in $\KE^n_B(G)$ has lower nesting depth than $A$ in $G$ since $B$ has been killed. Thus, we can apply the induction hypothesis to weaken each $A_i$ using $K_i$:
    \begin{align*}
      \sem{\KE^n_B(G) = \KE^n_B(H)} 
            & \Rightarrow\sem{\Wk{A_1}{K_1}\KE^n_B(G) = \Wk{A_1}{K_1}\KE^n_B(H)} \\
            & \ \ \vdots \\
            & \Rightarrow\sem{\Wk{A_n}{K_n}\ldots\Wk{A_1}{K_1}\KE^n_B(G) = \Wk{A_n}{K_n}\ldots\Wk{A_1}{K_1}\KE^n_B(H)}
    \end{align*}

    Invoking Lemma~\ref{lem:weaken_reorder} yields:
    \[ \sem{\KE^n_B (\Wk{A}{K}G) = \KE^n_B (\Wk{A}{K}H)} \]
    Again, since this holds for all $n$, we can conclude that all of the instances $\sem{\Wk{A}{K}G = \Wk{A}{K}H}$ hold, which completes the proof.
  \end{itemize}
\end{proof}

\begin{theorem}
  $(\Drop_B)$ is sound with respect to $\llbracket-\rrbracket$.
\end{theorem}
\begin{proof}
  For all $G, H$, $(\Drop_B)$ is derivable from existing rules. First, note that $\Kill_B\circ\Exp_{B,\fr}(G)$ is similar to $\Drop_B(G)$ except that some of the names have been replaced by fresh ones using $\fr$, which we can correct with suitable renaming functions $\rn_e, \rn_B$:
  \[
    \AxiomC{$G = H$}
    \RightLabel{\scriptsize($\Exp_{B}$)}
    \UnaryInfC{$\Exp_{B}(G) = \Exp_{B}(H)$}
    \RightLabel{\scriptsize($\Kill_B$)}
    \UnaryInfC{$\Kill_B\circ\Exp_{B}(G) = \Kill_B\circ\Exp_{B}(H)$}
    \RightLabel{\scriptsize(EdgeRename')}
    \UnaryInfC{$(\Kill_B\circ\Exp_{B}(G))[\rn_e] = (\Kill_B\circ\Exp_{B}(H))[\rn_e]$}
    \RightLabel{\scriptsize(BoxRename')}
    \UnaryInfC{$(\Kill_B\circ\Exp_{B}(G))[\rn_e][\rn_B] = (\Kill_B\circ\Exp_{B}(H))[\rn_e][\rn_B]$}
    \RightLabel{\scriptsize($\equiv$)}
    \UnaryInfC{$\Drop_B(G) = \Drop_B(H)$}
    \DisplayProof
  \]
\end{proof}

\fi


\end{document}